\documentclass[twocolumn]{aastex631}

\usepackage{bm}
\usepackage{amsmath}

\usepackage{xcolor}
\usepackage{textgreek}
\usepackage[utf8]{inputenc}
\usepackage[english]{babel}

\usepackage{hyperref}
\hypersetup{
    unicode, 
    colorlinks=true,
    linkcolor=linkcolor,
    citecolor=linkcolor,
    filecolor=linkcolor,
    urlcolor=linkcolor,
}
\usepackage{color,colortbl}
\definecolor{linkcolor}{rgb}{0.1,0.2,0.8}
\usepackage{tensind}
\tensordelimiter{?}
\DeclareGraphicsExtensions{.bmp,.png,.jpg,.pdf}
\usepackage{verbatim}
\usepackage[normalem]{ulem}
\usepackage{orcidlink}
\usepackage{soul}

\newcommand{\md}{\mathrm{d}}

\newcommand{\me}{\mathrm{e}}
\newcommand{\mi}{\mathrm{i}}

\newcommand{\bJ}{\bm{J}}

\newcommand{\br}{\bm{r}}

\newcommand{\btheta}{\bm{\theta}}
\newcommand{\bn}{\bm{n}}

\newcommand{\bOm}{\bm{\Omega}}
\newcommand{\nn}{\nonumber}

\newcommand{\p}{\partial}

\newcommand{\pattern}{\Omega_{\mathrm{p}}}

\newcommand{\JphivR}{$J_\varphi$-$\overline{v}_R$ }
\newcommand{\omegaeff}{\omega^\mathrm{eff}_{m,\pm}} 

\usepackage[normalem]{ulem}

\newcommand\CHrem{\bgroup\markoverwith{\textcolor{magenta}{\rule[0.5ex]{2pt}{0.9pt}}}\ULon}

\begin{document}

\title{On the radial velocity wave in the Galactic disk}

\author[0000-0002-5861-5687]{Chris Hamilton}
\affiliation{Institute for Advanced Study, Einstein Drive, Princeton NJ 08540}
\affiliation{Department of Astrophysical Sciences, Princeton University, 4 Ivy Lane, Princeton NJ 08544}

\author{Andrew Mummery}
\affiliation{Institute for Advanced Study, Einstein Drive, Princeton NJ 08540}

\author[0000-0002-8532-827X]{Joss Bland-Hawthorn}
\affiliation{Sydney Institute for Astronomy, School of Physics, A28, The University of Sydney, NSW 2006, Australia}

\correspondingauthor{Chris Hamilton} \email{chamilton@ias.edu}

\begin{abstract}
Stars in the Galactic disk have mean radial velocities $\overline{v}_R$ that oscillate as a function of angular momentum $J_\varphi$. This `\JphivR wave' signal also exhibits a systematic phase shift when stars are binned by their dynamical temperatures. However, the origin of the wave is unknown. Here we use linear perturbation theory to derive a simple analytic formula for the $J_\varphi$-$\overline{v}_R$ signal
that depends on the equilibrium properties of the Galaxy and the history of recent perturbations to it. The formula naturally explains the phase shift, but also predicts that different classes of perturbation should drive $J_\varphi$-$\overline{v}_R$ signals with very different morphologies.
Ignoring the self-gravity of disk fluctuations, it suggests that neither a distant tidal kick (e.g., from the Sgr dwarf) nor a rigidly-rotating Galactic bar can produce a qualitatively correct \JphivR signal. 
However, short-lived spiral arms can, and by performing an MCMC fit we identify a spiral perturbation that drives a \JphivR signal in reasonable agreement with the data.
We verify the analytic formula with test particle simulations, finding it to be highly accurate when applied to dynamically cold stellar populations. More work is needed to deal with hotter orbits, and to incorporate the fluctuations' self-gravity and the role of interstellar gas.
\end{abstract}



\section{Introduction}

The Gaia mission has revealed many interesting substructures in the chemo-dynamical phase space of our Galaxy --- see \cite{hunt2025milky} for a review. Key among these are the `radial velocity waves', namely oscillatory features that arise when the mean radial velocity $\overline{v}_R$ of a subset of stars within a given azimuthal range is plotted as a function of some variable like radius $R$, or guiding radius $R_\mathrm{g}$, or angular momentum $J_\varphi$.
Here we focus on the last of these --- angular momentum --- and hence upon what we call the `\JphivR wave' signal.

\begin{figure}
        \centering
        \includegraphics[width=0.99\linewidth]{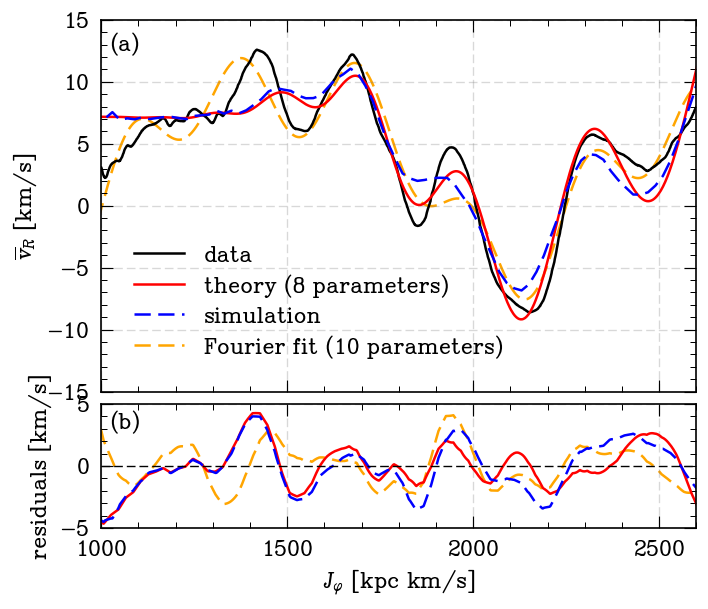}
        \caption{In panel (a), the black line shows the observed \JphivR signal in the Galactic disk for stars in an azimuthal wedge centered on the Sun ($\vert \varphi - \varphi_\odot\vert  < 0.2$ rad) and on dynamically cold orbits ($J_z < 3$ kpc km s$^{-1}$), following \cite{cao2024radial}.
        The red line shows our best analytic solution assuming a  transient spiral perturbation (Eq.~\eqref{eqn:vR_spiral}); the blue line shows the result of a test particle simulation with the same parameters. The gold dashed line is a ten-parameter fit to the data following Eq.~\eqref{eqn:friske}.
        See \S\ref{sec:fitting} for details.}
        \label{fig:best}
    \end{figure}
    
The \JphivR wave was first discovered by
\cite{friske2019more} using Gaia DR2 data, and has subsequently been revisited by various authors (e.g., \citealt{hunt2020power,antoja2022tidally,cao2024radial,khalil2025non,bernet2025dynamics}). In Fig.~\ref{eqn:friske}a, we show this signal extracted from the Gaia DR3 data following \cite{cao2024radial}, 
restricting to disk stars on rather cold orbits (vertical actions $J_z < 3$ kpc km s$^{-1}$) and within a narrow azimuthal wedge
centered on the Sun ($\vert \varphi - \varphi_\odot \vert < 0.2$ rad).
Visually it is clear that the \JphivR `wave' actually consists of multiple wave-like modes with comparable amplitudes, and wavelengths of several hundred kpc km s$^{-1}$ in angular momentum (equivalently, $\gtrsim 1$ kpc in guiding radius.) 
To be more precise, with a gold dashed line in Fig.~\ref{fig:best}a we plot a 
ten-parameter fit to the data in the range $J_\varphi \in (1000,2600)$ kpc km s$^{-1}$, assuming the functional form (see also \citealt{friske2019more}):
\begin{equation}
    \overline{v}_R = \sum_{j=1}^3 A_j\cos\left( \frac{2\pi J_\varphi}{\Lambda_j} + \chi_j \right) + u.
    \label{eqn:friske}
\end{equation}
The best fit returns
\begin{eqnarray}
    {\rm amplitudes,}\:\: \bm{A} &=& (3.03,6.78, 2.11)\: {\rm km\: s^{-1}}, \nonumber \\
    {\rm wavelengths,}\:\: \bm{\Lambda} &=& (308,1320,606)\: {\rm kpc\: km\: s^{-1}}, \nonumber \\
    {\rm phases,}\:\: \bm{\chi} &=& (3.26,5.61,0.45)\: {\rm rad}, \nonumber
\end{eqnarray}
and offset $u=3.85$ km s$^{-1}$. While the precise values do not matter much, it is useful to keep these ballpark numbers in mind throughout the paper.

The \JphivR wave is a clear sign of disequilibrium in the Milky Way's disk. The question is, what dynamical perturbation caused it? This has been the focus of various studies since \cite{friske2019more}'s original discovery. For instance, \cite{hunt2020power} found reasonable agreement between the \JphivR data and an N-body + hydrodynamical model of a bar-less, but otherwise Milky Way-like, 
isolated galaxy exhibiting transient spiral structure.
On the other hand, \cite{antoja2022tidally} and \cite{bernet2025dynamics} envisaged that the \JphivR signal is likely the imprint of a tidally-induced spiral, perhaps excited by the Sagittarius (Sgr) dwarf galaxy. However, as shown by \cite{bernet2025dynamics}, this type of perturbation tends to produce only two wave components in \JphivR space with very different amplitudes, rather than several comparable-amplitude waves that seem to make up the observed signal (Fig.~\ref{eqn:friske}). As such, these authors' models do not fit the data in a satisfying way.

Another attack on the problem was made by \cite{cao2024radial}, who paid particular attention to the phase shift $\Delta J_\varphi$ between \JphivR signals present in stellar populations with different dynamical temperatures (as originally reported in \citealt{friske2019more}). They tested whether this phase shift could be used to distinguish perturbations (bars, spirals, impulsive kicks, and combinations of these).  Though they gained some qualitative insights, 
they too were unable to identify a model that fit the data well.
Finally, 
while the \JphivR signal measures in-plane motions, another possibility is that it is related to a Sgr-induced corrugation wave (bending mode) that wraps up at about about half the angular rate of the tidally-induced spiral arms \citep{bland2021galactic}. However, a new analysis of the \cite{bland2021galactic} simulation shows that there is no clear correlation (Yamsiri et al 2026, MNRAS, submitted).

In this paper, we take a different approach to the \JphivR problem. The rationale is simple. The fluctuations involved are at the level of $\vert \overline{v}_R/V_0\vert \sim 5\%$ (where $V_0$ is the typical circular velocity), and from the typical wavelengths $\bm{\Lambda}$ embedded in the signal (Eq.~\eqref{eqn:friske}), one can deduce that the perturbation(s) responsible for it likely occurred only a handful of orbital periods ago (\S\ref{sec:interpretation}; see also \citealt{antoja2022tidally}). 
These properties mean the \JphivR wave problem is ideally suited to the application of linear perturbation theory \citep{Binney2018orbital,Monari2019-er,chiba2021resonance,petersen2024predicting,hamilton2024kinetic}.

The key result of this paper is shown by the red and blue curves in Fig.~\ref{fig:best}a. Our best fit to the analytic theory result $-$ shown in red $-$ is based on the linear response of an unperturbed disk to a single, two-armed transient spiral wave; in blue, we show the result of a test particle simulation we ran to confirm the theory. 
Although there are several caveats to this result, which we will discuss in detail below, 
it agrees with the \JphivR data better than the models listed above.
Moreover, Fig.~\ref{fig:best}b shows that it performs about as well as the theory-agnostic fit \eqref{eqn:friske}, despite having two fewer free parameters. 
The purpose of this paper is to present our analytic theory, to show how one can use it to gain insight into the \JphivR wave, to explain how it led us naturally to the result shown in Fig.~\ref{fig:best}, and to discuss what this may (or may not) teach us about Galactic structure.

The rest of this paper is organized as follows. In \S\ref{sec:theory} we write down the explicit analytic formula for the \JphivR wave driven by a prescribed perturbation (the detailed derivation is given in Appendix \ref{sec:derivation}).  
In \S\ref{sec:Examples} we give several concrete examples of relevant perturbations (Sgr, bar, and spiral) and discuss the \JphivR waves driven by each. We also verify numerically the regimes in which our formula is accurate.
In \S\ref{sec:fitting} we fit our analytic model to the Gaia data using MCMC sampling, identifying the transient spiral solution that gave rise to the red line in Fig.~\ref{fig:best}. We discuss the limitations and future extensions of our study in \S\ref{sec:Discussion}, and summarize in \S\ref{sec:Summary}.

\section{Theory}
\label{sec:theory}

We will use almost identical notation to that used in \cite{hamilton2026galactokinetics}.  
Thus, we restrict ourselves to two dimensions and a nearly-epicylic approximation for stellar orbits, mapping 
 from  position $\br = (\varphi, R)$ and velocity $\bm{v}=(v_\varphi, v_R)$ to angles $\btheta = (\theta_\varphi, \theta_R)$ and actions $\bJ = (J_\varphi, J_R)$.
 The azimuthal and radial frequencies are $\Omega_\varphi = \Omega + \Omega_\mathrm{D}$ and $\Omega_R = \kappa$, where $\Omega(J_\varphi)$ and $\kappa(J_\varphi)$ are the usual frequencies for circular orbits and $\Omega_\mathrm{D}(\bJ) \equiv \kappa'(J_\varphi)\, J_R < 0$, with the prime denoting a derivative with respect to $J_\varphi$,  is the Dehnen drift \citep{Grivnev1988,dehnen1999approximating,Binney2008-ou}.
 Although normally left out of the epicyclic approximation, this drift is often important in realistic calculations \citep{hamilton2026galactokinetics,GK2}, and  
 will contribute to the phase lag between hot and cold populations' \JphivR waves, as we will see. 
 Note that this coordinate mapping is accurate up to corrections $\mathcal{O}[(a_R/R_\mathrm{g})^2]$ with $a_R =(2J_R/\kappa)^{1/2}$ the epicyclic amplitude and $R_\mathrm{g}(J_\varphi)$ the guiding radius; typically these corrections are of order a few percent in the Milky Way disk(s) \citep{binney2023self}.
 We will have more to say about these approximations when our theory is tested numerically in \S\ref{sec:Examples}.

We define $\overline{v}_R$ as the mean value of $ v_R $ for a population of stars \textit{at a particular} $\varphi$ and $J_\varphi$\footnote{This is the only place our notation differs from that of \cite{hamilton2026galactokinetics}, whose overline denoted 
azimuthal averages over all $\varphi\in(0,2\pi)$.}; later, we will also average it over a narrow azimuthal wedge, consistent with the Gaia observations.
The derivation of our analytic formula for $\overline{v}_R$ is given in Appendix \ref{sec:derivation}. Briefly, the procedure is as follows. We multiply $v_R$ by the distribution function (DF) $f(\btheta, \bJ, t)$ of stellar orbits and perform the appropriate phase-space integrals. Those integrals pick out particular Fourier components of the DF, and it is these that we compute using linear response theory, under the assumption 
that the
relevant potential perturbations are sufficiently weak ($\sim \eta V_0^2$ where $\eta \ll 1$).
To make the linear response calculation analytically tractable, we further assume that  the initial DF has  a Schwarzchild form \citep{binney2023self}
\begin{equation}
f_0(\bJ) = \frac{1}{2\pi\langle J_R \rangle} F_0(J_\varphi) \me^{-J_R/\langle J_R \rangle},
\label{eqn:Schwarzchild}
\end{equation}
where $\langle J_R\rangle$ can in principle depend on $J_\varphi$, and that $\epsilon \equiv (\langle J_R \rangle/J_\varphi)^{1/2} \ll 1$ is sufficiently small that we can use the long wavelength approximation from \cite{hamilton2026galactokinetics}.
With these assumptions we arrive at Eq.~\eqref{eqn:meanvr2yet}.
Finally, we focus on timescales $t \lesssim 1$ Gyr (to be justified below), which allows us to make one last simplification \eqref{eqn:Dehneneasier}.
The final result is 
\begin{align}
    &\overline{v}_R(\varphi, J_\varphi, t) \simeq \mathrm{Re} \sum_{m=1}^\infty 
    \sum_{\pm}  \me^{i(m\varphi-\omega^\mathrm{eff}_{m,\pm} t)} 
    \nn
    \\
   &\,\,\,\,\,\,\,\,\,\,\times \int_0^t \md t'  \me^{i \omega^\mathrm{eff}_{m,\pm} t'}  \delta\Phi_{m}(t')  \left[\pm \frac{\gamma m}{R_\mathrm{g}} - \mi k_R(t')\right].
   \label{eqn:meanvr2}
\end{align}
 Here $\gamma \equiv 2\Omega/\kappa$, while $\delta \Phi_{m}$ is the $m$th azimuthal Fourier component of the potential perturbation $\delta \phi$ evaluated at guiding radius $R_\mathrm{g}$ and time $t'$, and $k_R$ is its (generally complex) radial wavenumber (defined in Eq.~(64) of \citealt{hamilton2026galactokinetics}). We also defined the effective frequency
\begin{equation}
    \omegaeff(J_\varphi, \langle J_R \rangle) \equiv m(\Omega + 2 \kappa' \langle J_R \rangle ) \pm \kappa.
    \label{eqn:omegaeff}
\end{equation}

Equations \eqref{eqn:meanvr2}-\eqref{eqn:omegaeff} constitute the key analytic result of this paper.
Importantly, they do not depend on $F_0(J_\varphi)$.

\subsection{Interpretation}
\label{sec:interpretation}

To evaluate Eqs.~\eqref{eqn:meanvr2}-\eqref{eqn:omegaeff} explicitly,
we need to stipulate the potential perturbations $\delta \phi(\br, t)$ experienced by the stars. We will do this for several examples in \S\ref{sec:Examples}.
However, we can already make a few general remarks that should hold for any realistic perturbation, as long as the approximations used to derive \eqref{eqn:meanvr2}-\eqref{eqn:omegaeff} are satisfied.

Naively, the first line in \eqref{eqn:meanvr2} tells us that a perturbation with $m$-fold azimuthal symmetry wants to launch two distinct radial velocity signals in $J_\varphi$ space, which rotate in azimuth with frequencies 
$\omega^\mathrm{eff}_{m,\pm}$ (Eq.~\eqref{eqn:omegaeff}).  
If we focus on a fixed $\varphi$, then since $\omegaeff$ are monotonic functions of $J_\varphi$, 
for $\omegaeff  t \gtrsim 1$, these signals will resemble waves in $J_\varphi$ space with approximate `wavelength'
\begin{align}
 & \Lambda_m^\pm   =  2\pi \bigg\vert \frac{\md \omega^\mathrm{eff}_{m,\pm}}{\md J_\varphi} t \bigg\vert^{-1} \label{eqn:wavelength_Jphi},
    \\
       &\simeq  
              \begin{pmatrix}
           260 \,\mathrm{kpc \, \mathrm{km} \,\mathrm{s}^{-1}} \\ 1500 \,\mathrm{kpc \, \mathrm{km} \,\mathrm{s}^{-1}}
       \end{pmatrix} 
       \times    \frac{ J_\varphi}{1760\mathrm{\,kpc\,km\: s^{-1}} } \left( \frac{t/T_\varphi}{0.5} \right)^{-1} ,
       \label{eqn:wavelength_estimate}
\end{align}
where $T_\varphi = 2\pi/\Omega$. (In the second line we assumed $m=2$ and a flat rotation curve, so $\gamma=\sqrt{2}$ --- we will make these assumptions whenever writing down numerical estimates throughout the paper.)
Thus, the wavelength of the $J_\varphi$-$\overline{v}_R$ signal increases with $J_\varphi$ and decreases with time.
In particular, by comparing the estimate \eqref{eqn:wavelength_estimate} to the wavelength values inferred using Eq.~\eqref{eqn:friske}, 
we surmise that the \JphivR signal is likely only a few orbits old (see also \citealt{antoja2022tidally}).
This justifies a posteriori our use of the simplification \eqref{eqn:Dehneneasier} when deriving Eq.~\eqref{eqn:meanvr2}.

However, this naive picture is complicated by the second line in \eqref{eqn:meanvr2}, which tells us that the  signal
is in fact modulated by the
prior history of perturbations from all $t'<t$, as well as the factor  $\me^{i \omega^\mathrm{eff}_{m,\pm} t'}$ that picks out any resonant driving. The amplitude and phase of each signal depends crucially on the perturbation's azimuthal ($m/R_\mathrm{g}$) and radial ($k_R$) wavenumbers, the latter of which can also be time-dependent. So, while we expect the naive interpretation just given to be accurate for very short-lived perturbations, things can be somewhat more complicated for longer-lived perturbations, as we will see.

Next, the term $2\kappa'\langle J_R \rangle$ in \eqref{eqn:omegaeff} accounts for the fact that, at a fixed angular momentum, eccentric orbits drift in azimuth relative to circular ones. This will lead to a phase shift in the \JphivR  signals between dynamically hot and cold stellar populations\footnote{\cite{cao2024radial} attributed the phase shift to the dependence of \textit{radial} frequency on dynamical hotness; however, the drift correction to the azimuthal frequency $\Omega_\mathrm{D}$ is the dominant effect \citep{dehnen1999approximating,hamilton2026galactokinetics}.}. In fact, in the case of a very short-lived perturbation, where there is no modulation of the signals arising from the second line in \eqref{eqn:meanvr2}, this phase shift can be calculated precisely. To do this, suppose we had two distinct stellar populations, one comprising purely circular orbits and one with a nonzero $\langle J_R \rangle$. 
We equate the phase $m\varphi - \omegaeff t$ for the two populations at a fixed azimuth and time, but allow the latter population to be shifted by $\Delta J_\varphi^\pm$:
\begin{align}
    &\omega^\mathrm{eff}_{m,\pm}(J_\varphi, 0)  = \omega^\mathrm{eff}_{m,\pm}(J_\varphi + \Delta J_\varphi^\pm, \langle J_R \rangle).
\end{align}
Expanding the right-hand side for small $\Delta J_\varphi$, using 
\eqref{eqn:omegaeff}, and keeping only the lowest-order surviving terms, 
we soon arrive at
\begin{align}
    \Delta J_\varphi^\pm &\simeq - \frac{2m\kappa'}{m\Omega'\pm\kappa'}\langle J_R\rangle
    \label{eqn:phase_shift}
    \\
       &\simeq  
              \begin{pmatrix}
           -33 \,\mathrm{kpc \, \mathrm{km} \,\mathrm{s}^{-1}} \\ -193 \,\mathrm{kpc \, \mathrm{km} \,\mathrm{s}^{-1}}
       \end{pmatrix}     \times   \frac{\langle J_R \rangle}{20\mathrm{\,kpc\,km\: s^{-1}} }.
\end{align}
The overall minus sign in this result means the $J_\varphi$-$\overline{v}_R$ signal of a hot population will generically \textit{lag} that of a cold population \citep{cao2024radial}.
We emphasize that this result is only precise for very short-lived kicks;
for longer lived perturbations like bars and spirals, it is only a rough estimate, but the physical picture behind the phase offset remains valid.

We have not had to stipulate here whether the potential perturbation $\delta \phi(\br, t)$ includes the self-gravity of the DF fluctuations or is 
an externally-applied field. Thus, we expect these conclusions to hold even for self-consistent calculations provided we stipulate $\delta \phi$ correctly.

\section{Examples}
\label{sec:Examples}

In this section we illustrate the behavior of the \JphivR signal for several example perturbations using our analytic formulae \eqref{eqn:meanvr2}-\eqref{eqn:omegaeff}. We also check the accuracy of our analytic results using numerical simulations. 
We ignore the role of self-gravity for these examples, treating our stars as test particles subject to externally-prescribed forces (but see the Discussion).

\subsection{Distant impulsive kick (Sgr dwarf)}
\label{sec:impulse}

We first consider an idealized version of an impulsive kick from a distant perturber like the Sgr dwarf, thereby recovering (and generalizing) some results of \cite{antoja2022tidally,bernet2025dynamics}. Following those authors, we write 
\begin{equation}
    \delta \phi(\br, t) = -\eta \,\frac{V_0^2}{2} \, T \delta (t) \bigg( \frac{R}{R_\mathrm{ref}}\bigg)^2 \cos(2\varphi),
\label{eqn:deltaphi_impulse}
\end{equation}
where $\eta$ is a dimensionless perturbation strength,  $V_0$ has units of velocity, $T$ has units of time, and $R_\mathrm{ref}$ is some reference radius. 
Given this choice of potential perturbation, the only azimuthal Fourier component we need consider is $m=2$; the corresponding radial wavenumber is $k_R = -{2i}/{R_\mathrm{g}}$.
Plugging into \eqref{eqn:meanvr2}, we get 
\begin{equation}
    \overline{v}_R \simeq - \eta \frac{V_0^2}{2} T  \frac{R_\mathrm{g}}{R_\mathrm{ref}^2} \sum_{\pm}(\pm \gamma-1) \cos(2\varphi - \omega^\mathrm{eff}_{2,\pm}t).
    \label{eqn:vR_analytic_impulse}
\end{equation}
This is clearly a sum of two waves with frequencies $\omega^\mathrm{eff}_{2,\pm}$. If the amplitudes of these waves are $A_\pm$, then they are in the ratio 
\begin{equation}
    \bigg\vert \frac{A_-}{A_+}\bigg\vert = \frac{\gamma+1}{\gamma-1} \simeq 5.8.
    \label{eqn:amplitude_ratio_impulse}
\end{equation}
This confirms the finding of \cite{bernet2025dynamics}: a distant tidal kick overwhelmingly excites the `slow' ($-$) wave compared to the `fast' ($+$) wave. In fact, our result is slightly more general than that of \cite{bernet2025dynamics} because we account for a non-zero velocity dispersion (so our waves have frequency $\omegaeff$ rather than $m\Omega \pm \kappa$).

\begin{figure*}
    \centering
    \includegraphics[width=0.95\linewidth]{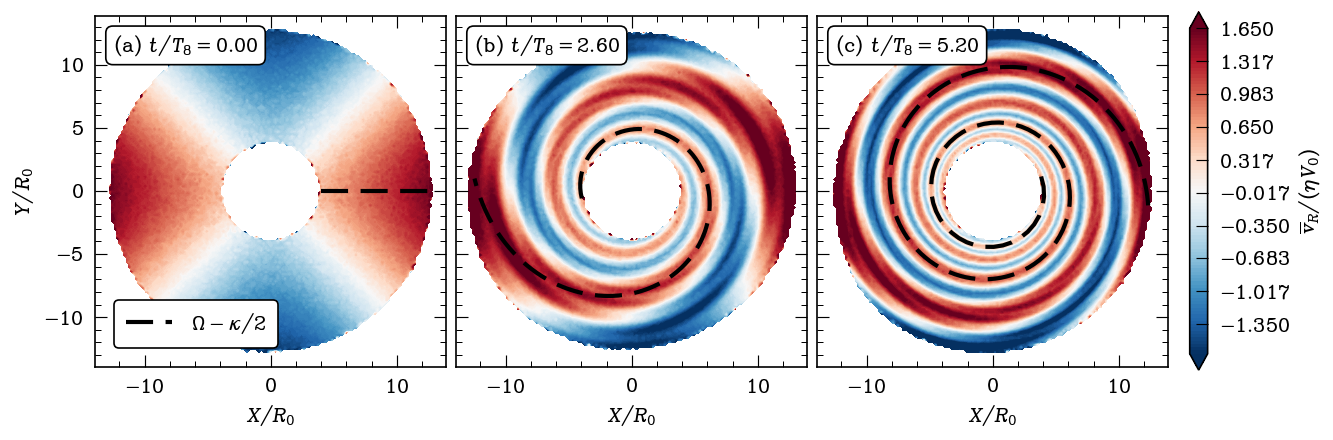}
    \caption{Radial velocity signal for a dynamically very cold ($\epsilon=0.01$) population of $N=10^7$ test particles subjected to the impulsive quadrupolar perturbation \eqref{eqn:deltaphi_impulse}. Colors show contours of the mean radial velocity $\overline{v}_R$ in real space $(X,Y)=(R\cos\varphi, R \sin \varphi)$ at three different times $t/T_8$, extracted from test particle simulations. The dashed lines in these panels show the loci of the curves $\varphi = (\Omega-\kappa/2) t$ respectively, where the frequencies are evaluated at the guiding radius $R_\mathrm{g}=R$. }
    \label{fig:impulsive_cold_contours}
\end{figure*}
\begin{figure}
    \centering
    \includegraphics[width=0.95\linewidth]{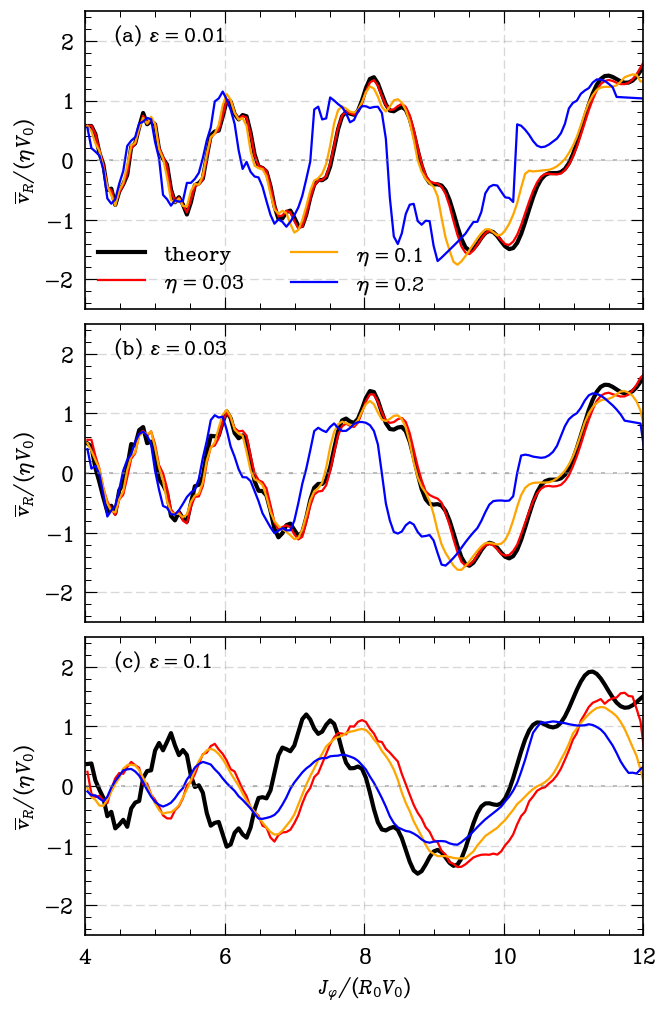}
    \caption{The \JphivR signal (averaged over $\vert \varphi \vert < 0.2$) generated by the quadrupolar impulsive kick \eqref{eqn:deltaphi_impulse}, measured at time $t=5.2 T_8$, for various values of $\eta$ and $\epsilon$. Colored lines show the results of $N=5\times 10^7$ test particle simulations, while black lines show the result of analytic theory (Eq.~\eqref{eqn:vR_analytic_impulse}).}
    \label{fig:linearity_test}
\end{figure}

To test the result in \eqref{eqn:vR_analytic_impulse}, consider a galaxy with a flat rotation curve with circular speed
$V_0$, and subject it to the perturbation \eqref{eqn:deltaphi_impulse} with $R_\mathrm{ref}=8R_0$ and 
$T=0.16T_8$  where $T_8 \equiv 2\pi\times 8R_0/V_0$. We study the motion of test particles in the midplane of this galaxy whose initial conditions are drawn from the Schwarzchild DF in Eq.~\eqref{eqn:Schwarzchild}. For simplicity, we fix the mean radial action to be $\langle J_R \rangle = \epsilon^2 \times 8R_0V_0$ at all $J_\varphi$. For a given $\epsilon$, this corresponds to a radial velocity dispersion at radius $R$ given by
\begin{equation}
    \sigma \simeq 26 \,\mathrm{km\: s^{-1}} \times \left(\frac{\epsilon}{0.1}\right)\left( \frac{R}{8\,\mathrm{kpc}}\right)^{-1/2}\left( \frac{V_0}{220 \,\mathrm{km\: s^{-1}}}\right).
    \label{eqn:sigma}
\end{equation}
Since \eqref{eqn:meanvr2} is independent of $F_0$, we do not choose a physically-motivated (e.g., exponentially decreasing) $F_0$, but rather one proportional to $J_\varphi$ in order to suppress Poisson noise at larger angular momenta.\footnote{In particular this choice means that our $F_0$ cuts off unphysically at some $J_\varphi^\mathrm{min/max}$. In principle this could make terms $\propto \p f_0/\p J_\varphi$, which we ignored in our theory (see Appendix \ref{sec:derivation}), more important. We never find this to be a significant issue as long as we take $J_\varphi^\mathrm{min/max}$ well outside the range of interest, so we cease to mention this subtlety from now on.}

With this setup, to define the system, we need only stipulate the dimensionless parameters $\eta$ and $\epsilon$.
Having chosen these, we may calculate the right-hand side of the analytic linear theory formula \eqref{eqn:vR_analytic_impulse}. We can also run a test particle simulation of the same system, drawing $N$ particles in the range $J_\varphi/(R_0V_0)\in(4,12)$ and integrating their orbits using \texttt{galpy} \citep{bovy2015galpy}.

First we consider a very cold population of $N=10^7$ stars with $\epsilon = 0.01$, and a kick strength $\eta = 0.1$.
In Fig.~\ref{fig:impulsive_cold_contours}, we show contours of $\overline{v}_R/(\eta V_0)$ in real space
(cf. \citealt{bernet2025dynamics}) extracted from the simulation at different times $t/T_8$.
These contours form a quadrupolar structure that winds differentially into a spiral. 
The dominant signal wraps with a pattern speed very close to $\Omega - \kappa/2$, as expected from Eq.~\eqref{eqn:amplitude_ratio_impulse}.
This behavior is entirely generic, although with dynamically hotter populations, the spiral features become less well defined due to epicyclic blurring. 

We now turn to the \JphivR signal proper, which for the rest of this section we compute by averaging over the azimuthal range $\vert \varphi\vert < 0.2$. We also increase the number of particles to $N=5\times 10^7$ to improve numerical resolution. In Fig.~\ref{fig:linearity_test}, we show a selection of these signals --- again normalized by $\eta V_0$ --- for 
different combinations $\epsilon, \eta$, all at time $t/T_8 = 5.2$. Different colored lines correspond to simulation results while the thick black lines show the linear theory result \eqref{eqn:vR_analytic_impulse} averaged over $\vert \varphi \vert < 0.2$.

From panels (a) and (b) of Fig.~\ref{fig:linearity_test}, we see that at the lowest values of $\epsilon \lesssim 0.03$,
our analytic theory works extremely well for small $\eta$ values. As $\eta$ is increased to $\gtrsim 0.1$, the analytic theory starts to break down, because the perturbation can no longer be considered linear. 
On the other hand, for $\epsilon= 0.1$ (panel (c)), our analytic prediction is never perfectly accurate for \textit{any} $\eta$. This is because of the breakdown of the long-wavelength assumption which was used to simplify the linear calculation (Appendix \ref{sec:derivation}). More precisely, the more eccentric an orbit is, the more of the spatial structure of the perturbing potential it explores, and the more Fourier components need to be included in the linear theory, whereas the long wavelength assumption retains only two of these components. (Note that the low-$\eta$ (red and orange) curves do still agree well in this panel, confirming that it really is the long-wavelength approximation, rather than the linear approximation, that is causing the theory to break down.)
These deviations are important because realistic disk populations often have $\epsilon \gtrsim 0.1$ (cf. Eq.~\eqref{eqn:sigma}). 
In any case, the qualitative behavior predicted by the analytic theory still holds: we have one dominant wave, its wavelength increases with $J_\varphi$, etc.

The results of this subsection suggest that a single, impulsive, quadrupolar kick --- \textit{without} any self-gravitating collective response, as we are assuming here --- is unlikely to be responsible for the \JphivR signal in our Galaxy. The key reason is that the signal produced by such a kick is dominated by a single sinusoidal component (Eqs.~\eqref{eqn:vR_analytic_impulse}-\eqref{eqn:amplitude_ratio_impulse}), rather than multiple components of comparable amplitude as is observed (Eq.~\eqref{eqn:friske}).

\subsection{Rigidly-rotating bar}
\label{sec:bar}
Next we consider the response to a rotating bar. 
At the radii of interest (far outside the bar radius),  a good model is \citep{dehnen2000effect,chiba2021resonance}:
\begin{align}
    \delta \phi(\bm{r}, t) =&   -\eta \frac{V_0^2}{2}\left(\frac{R}{R_\mathrm{b}}\right)^{-3} \cos[2(\varphi-\pattern t)] B(t),
    \label{eqn:phi_bar}
\end{align}   
    where $R_\mathrm{b}$ is the bar radius and $\pattern$ is its pattern speed.
    The function $B(t)$ tells us how rapidly the bar potential is `switched on'. For simplicity, we will just look at two special cases, one in which the bar is switched on instantaneously at $t=0$ (so $B=1$ always), and one in which it is turned on adiabatically ($B=\me^{\beta t}$ where $\beta \to 0^+$ for $t<0$, and $B=1$ for $t>0$). Neither case is particularly realistic --- bars grow neither instantaneously nor adiabatically, and in the latter case we should not really be using linear theory anyway (e.g., \citealt{Chiba2022-qt}).   Nevertheless, in these limits, we can derive explicit analytic results that are useful for interpreting more realistic calculations.

Just as in \S\ref{sec:impulse}, the only nonzero Fourier component of the potential perturbation
is for $m=2$,  and again the radial wavenumber turns out to be purely imaginary, $k_R=3i/R_\mathrm{g}$. From Eq.~\eqref{eqn:meanvr2}, we find
\begin{align}
    \overline{v}_R \simeq -& \eta \frac{V_0^2 }{2} \frac{R_\mathrm{b}^3}{R_\mathrm{g}^4} \sum_{\pm } \Big( \pm \gamma + \frac{3}{2} \Big)  \nn 
   \\ \,\,\, &\times  
    \frac{\sin(2\varphi - 2\pattern t) - s\sin(2 \varphi - \omega^\mathrm{eff}_{2,\pm}t)}{\omega^\mathrm{eff}_{2,\pm} - 2\pattern},
    \label{eqn:meanvrbar}
\end{align}
where $s=1$ ($s=0$) for switching on instantaneously (adiabatically).

\begin{figure}
    \centering
    \includegraphics[width=0.995\linewidth]{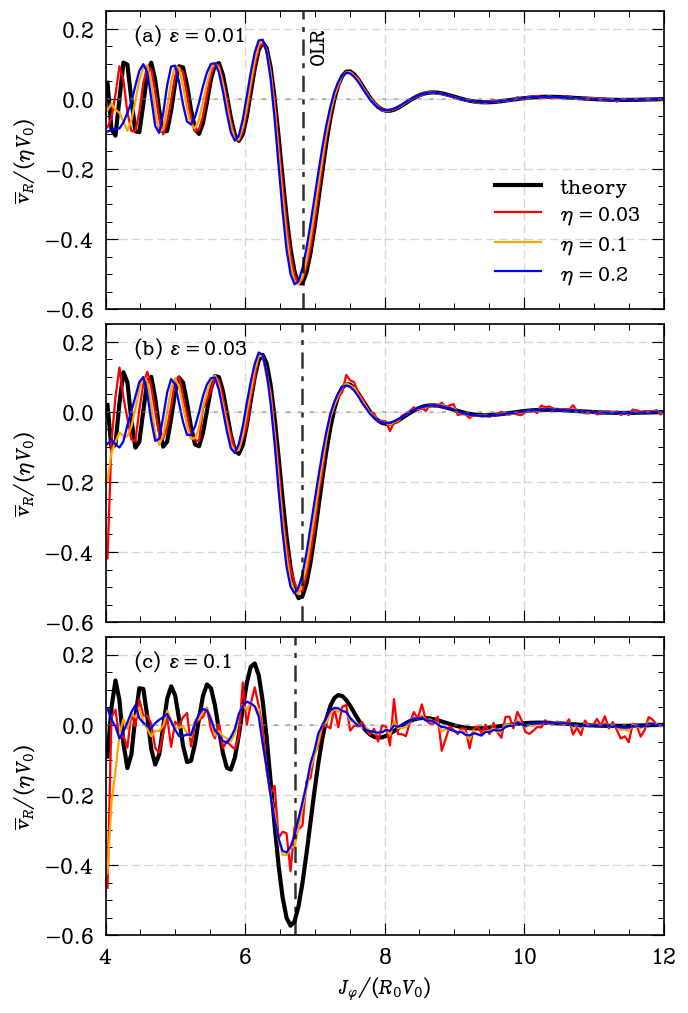}
    \caption{Similar to Fig.~\ref{fig:linearity_test}, except 
    using the rigidly-rotating bar potential \eqref{eqn:phi_bar}, with $R_\mathrm{b}=2R_0$, $R_\mathrm{CR}=4R_0$, and $s=1$ (instantaneous switch on). We run the simulation to $t=2T_8$. The OLR location $\omega_{2,-}^\mathrm{eff} = 2\pattern$ is shown with a vertical dot-dashed line in each panel.}
    \label{fig:bar_lines}
\end{figure}

Consider  the instantaneous switch-on limit ($s=1$). In this case, Eq.~\eqref{eqn:meanvrbar} tells us that the radial velocity signal consists of three wave-like components; two with the usual frequencies $\omega_{2,\pm}^\mathrm{eff}$, and one tied to the pattern speed of the bar, i.e., with frequency $2\pattern$.
The denominator in \eqref{eqn:meanvrbar} tells us that the combined \JphivR response of these waves will have a peak near the outer Lindblad resonance (OLR)\footnote{The inner Lindblad resonance (ILR) $\omega_{2,-}^\mathrm{eff} = 2\pattern$ tends to sit inside the bar radius, so is ignorable for our purposes.} $\omega_{2,+}^\mathrm{eff} = 2\pattern$. This may seem a promising way to reproduce the 
observed  morphology in Fig.~\ref{fig:best} (e.g., by placing the bar's OLR near $J_\varphi = 2000$ kpc km s$^{-1}$).
However, in the end this idea seems to fail. The reason is that the amplitude ratio of the two waves associated with $\omega_{2,\pm}^\mathrm{eff}$ is bounded by
\begin{equation}
    \bigg\vert \frac{A_-}{A_+}\bigg\vert \lesssim \bigg\vert  \frac{3-2\gamma}{3+2\gamma} \bigg\vert  \simeq 0.03.
    \label{eqn:amplitude_ratio_bar}
\end{equation}
This means the signal will be dominated by just one or two waves, rather than three as seems to be the case in the observed pattern (Eq.~\eqref{eqn:friske} and \S\ref{sec:fitting}).

To illustrate this explicitly, we present Fig.~\ref{fig:bar_lines}. We made this plot using the same procedure as for Fig.~\ref{fig:linearity_test}, except this time we perturbed the system with the bar potential \eqref{eqn:phi_bar}, choosing $R_\mathrm{b}=2R_0$, $R_\mathrm{CR}=4R_0$, and $s=1$, and measured the \JphivR wave at $t/T_8 = 2.0$.
We see that the linear theory expression \eqref{eqn:meanvrbar} (averaged over $\vert \varphi \vert < 0.2)$ again does an excellent job of predicting the \JphivR wave behavior for $\epsilon \lesssim 0.03$ (panels (a) and (b) of Fig.~\ref{fig:bar_lines}), with only a slight phase offset creeping in at the higher $\eta$ values due to nonlinearity. For $\epsilon = 0.1$ (panel (c)), the theory is again qualitatively accurate but quantitatively it over-predicts the amplitude of the signal at all $\eta$. As in Fig.~\ref{fig:linearity_test}, this discrepancy is mostly due to the fact that we used the long-wavelength approximation to neglect various terms in the linear theory.\footnote{The additional jaggedness of the red line in panel (c) is due to finite-$N$ noise, which can be a significant fraction of the signal when the perturbation strength $\eta$ is very small.}

Regardless of the accuracy of the analytic theory, in all cases in Fig.~\ref{fig:bar_lines} we see the expected peak in the \JphivR signal near the OLR (cf. Eq.~\eqref{eqn:meanvrbar}); this is reminiscent of the bottom-left panel of \cite{cao2024radial}'s Fig.~8, which also showed the test particle response to a bar perturbation. However, away from this resonant peak the wave is dominated by a single sinusoidal component. It does not exhibit any of the intricate beating structure that arises when one has several equal-amplitude waves overlapping, as we see in the data (Fig.~\ref{fig:best}). The signal in Fig.~\ref{fig:bar_lines} also suffers from the $\propto R_\mathrm{g}^{-3}$ decay inherent in the bar potential \eqref{eqn:phi_bar} which means it is very weak at radii beyond the OLR.  Though we tried tweaking the bar parameters we were unable to create any \JphivR signal that resembled the data in Fig.~\ref{fig:best}.

If instead we consider the adiabatic case ($s=0$ in Eq.~\eqref{eqn:meanvrbar}),
we find that the only surviving wave is that tied to the bar pattern speed. 
Physically, this is because the rest of the signal has phase-mixed away in the infinite time taken for the bar to switch on.  The response at the outer Lindblad resonance in this case actually diverges, which reflects the fact that our linear theory breaks down here and should not really be trusted.
Nevertheless, the lack of multiple wave components in this limit suggests that a slowly-grown bar will also fail to produce a \JphivR wave consistent with observations.

We do not show here the case here of a bar with a decreasing pattern speed \citep{chiba2021resonance}.
However, given the wavelength of the observed \JphivR signal, it seems likely that it is at most a few dynamical times old (see Eq.~\eqref{eqn:wavelength_Jphi}), and the pattern speed of our Galaxy's bar cannot have changed by much over that timescale. Though we do not show it here, we did perform a perturbative calculation in which $\pattern$ was allowed to change fractionally, but the results did not alter any of our qualitative conclusions. In summary, it seems that the Galactic bar alone --- in the test particle limit, \textit{without} an accompanying collective self-gravitating response --- is unlikely to be responsible for the observed \JphivR wave.

\subsection{Transient spiral}
\label{sec:spiral}

We now consider the \JphivR waves driven by a transient spiral perturbation.
We consider a nearly-logarithmic spiral,
multiplied by Gaussians in both radius and time:
\begin{align}
&\delta \phi(\br, t) \equiv -\eta \frac{V_0^2}{2}\sqrt{\frac{R_\mathrm{ref}}{R}}  \me^{-(R-R_\mathrm{CR})^2/(2R_\beta^2)} \me^{-t^2/(2\tau^2)} \nn
    \\
    &\,\,\,\,\,\,\,\times \cos[m(\varphi-\varphi_0-\pattern t)+m\cot\alpha\,\ln (R/R_\mathrm{ref})].
    \label{eq:spiral_potential}
\end{align}
Here $\tau$ sets the spiral's lifetime, $\alpha$ is its pitch angle (assumed constant for now), and $\varphi_0$ is a constant phase offset. The Gaussian in $(R-R_\mathrm{CR})$ accounts for the fact that spirals tend to have finite radial envelopes that peak in the vicinity of corotation. 
The parameter $R_\beta \equiv \beta \sqrt{2} R_\mathrm{CR}/m$ sets the width of this radial envelope, and for a flat rotation curve is equal to $\beta$ times the distance between the corotation resonance and the Lindblad resonances. 
Had we not included the Gaussian in $(R-R_\mathrm{CR})$, the potential perturbation \eqref{eq:spiral_potential} would have had an envelope decaying as $ \propto R^{-1/2}$, and would have corresponded self-consistently to a nearly-logarithmic surface density perturbation with an envelope $\propto R^{-3/2}$  (see \S2.6.3 of \citealt{Binney2008-ou}). 
But with the Gaussian factor included, it is not guaranteed that the corresponding surface density $\delta \Sigma$ that generated it is physically sensible (e.g., does not lead to negative or divergent overall densities $\Sigma = \Sigma_0 +\delta \Sigma$).
Instead, this should be checked for each set of spiral parameters.  While we don't perform this check explicitly in this section (since our aim is illustration, not realism), we do when fitting the data in \S\ref{sec:fitting}.

 We now plug this choice of potential perturbation into Eq.~\eqref{eqn:meanvr2}.
 We also  shift the origin of time from $t=0$ to $t=-\infty$ in Eq.~\eqref{eqn:meanvr2}, to accommodate the fact that our spiral peaks at $t=0$. After some algebra, the result is
\begin{align}
    \overline{v}_R &\simeq -\frac{1}{2}\sqrt{\frac{\pi}{2}} \eta \tau \frac{V_0^2}{2} \sqrt{\frac{R_\mathrm{ref}}{R_\mathrm{g}}}
    \me^{-(R_\mathrm{g}-R_\mathrm{CR})^2/(2R_\beta^2)}
    \nn
    \\
&    \times  \mathrm{Re}
    \sum_{\pm} \me^{i[m(\varphi - \varphi_0) - \omegaeff t + m\cot \alpha \ln (R/R_\mathrm{ref})]} 
    \nn
    \\
    & \times   
\me^{- (\omega^\mathrm{rel}_{m,\pm} \tau)^2/2}\Big[1 + \mathrm{erf}\Big( \frac{t - i \omega^\mathrm{rel}_{m,\pm} \tau^2 }{\sqrt{2}\tau}\Big) \Big] \nn
\\
&\times \Big[\pm \frac{m\gamma}{R_\mathrm{g}} + \frac{1}{2R_\mathrm{g}} + \frac{R_\mathrm{g}-R_\mathrm{CR}}{2R_\beta^2} - \frac{im\cot\alpha}{R_\mathrm{g}}\Big],
    \label{eqn:vR_spiral}
\end{align}
where
\begin{equation}
    \omega^\mathrm{rel}_{m,\pm} \equiv \omegaeff - m\pattern.
\end{equation}

\begin{figure*}
    \centering
    \includegraphics[width=0.95\linewidth]{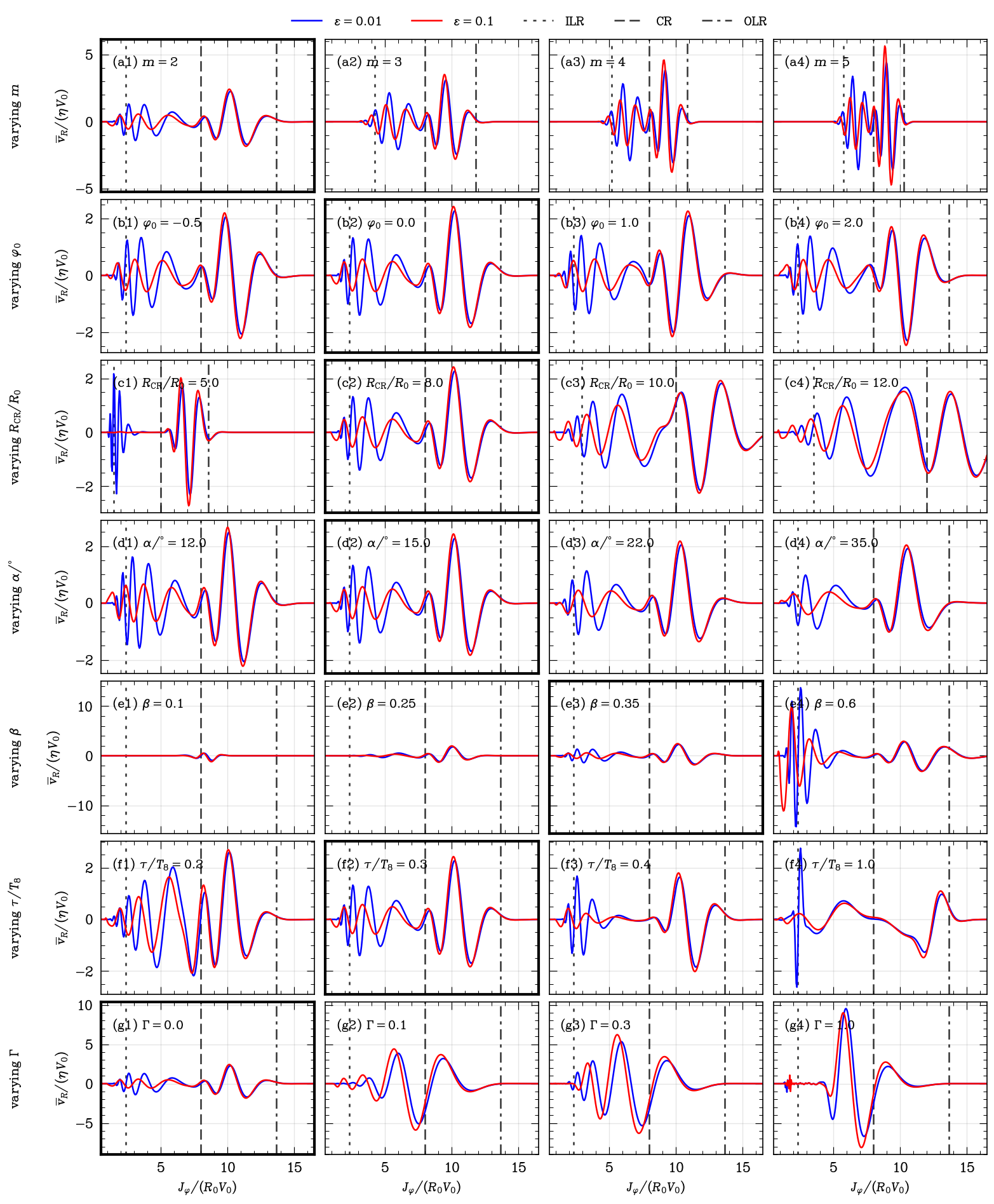}
    \caption{We plot the formula \eqref{eqn:vR_spiral}, averaged over the azimuthal range $\vert \varphi \vert < 0.2$.
    The fiducial parameters for rows (a)-(f) are $m=2$, $\varphi_0=0$, $R_\mathrm{CR}=R_\mathrm{ref}=8R_0$, $\alpha = 15^\circ$, $\beta=0.35$, $\tau/T_8=0.3$ and $t/T_8=1.0$. In row (g) 
    we perform the same calculation except for a shearing spiral with shear rate $\Gamma$ (see Eq.~\eqref{eqn:shearing_wave}). The $\Gamma=0$ case is equivalent to the fiducial case in the rows above.  
    In each row we vary only a single parameter compared to the fiducial model, as indicated in the panel labels, and in each panel we plot two curves, one for $\epsilon=0.01$ (blue) and another for $\epsilon = 0.1$ (red).
    The fiducial panel is reproduced once in each row with a bold frame (but note each row has a unique vertical axis). Resonance locations for circular orbits are shown with black vertical lines.}
    \label{fig:spiral_grid}
\end{figure*}

Although the result in Eq.~\eqref{eqn:vR_spiral} looks complicated, each part of it has a simple physical interpretation; we illustrate this using Fig.~\ref{fig:spiral_grid}. Rows (a)-(f) of this figure show the right-hand side of \eqref{eqn:vR_spiral} averaged over the azimuthal range $\vert \varphi \vert < 0.2$ for both $\epsilon=0.01$ (blue) and $\epsilon = 0.1$ (red), using the same flat rotation curve and initial DF in \S\S\ref{sec:impulse}-\ref{sec:bar}, now measuring the signal at $t/T_8 = 1.0$. (We also 
extend the $J_\varphi$ range used compared to previous plots, although one should be careful that our nearly-epicyclic approximation will certainly fail at small enough angular momenta.)
In each row, we vary only a single parameter compared to the fiducial model, whose values are given in the figure caption.
Panel (g1) of Fig.~\ref{fig:spiral_grid} again shows the fiducial result, but in panels (g2)-(g4)
we show the result of a different calculation where we replace the rigidly-rotating spiral of Eq.~\eqref{eq:spiral_potential} with a shearing one, via the substitutions 
\begin{equation}
   \pattern \to 0\,, \,\,\,\,\,\,\,\,\,\,\,\,  \cot \alpha \to \cot\alpha + \Gamma \Omega(R) t,
    \label{eqn:shearing_wave}
\end{equation}
for some dimensionless shear rate $\Gamma$.  For the flat rotation curve we are assuming here, $\Gamma=1.0$ corresponds to a winding material arm, while $\Gamma = 1-1/\sqrt{2} \simeq 0.3$ corresponds to a Lindblad-Kalnajs kinematic density wave \citep{Binney2008-ou,bland2021galactic,GK2}.\footnote{For the shearing spirals, since there is no overall pattern speed, there are strictly no resonances any more.
However, $R_\mathrm{CR}$ still has meaning as the place where the radial envelope peaks
For instance, we can approximate the ``shearing sheet'' model of swing amplification \citep{julian1966non,binney2020shearing} by setting $\Gamma=1$ and placing $R_\mathrm{CR}$ at the center of the sheet. For illustrative purposes we still plot the resonance lines in row (g) of Fig.~\ref{fig:spiral_grid} as if we had set $\pattern=V_0/R_\mathrm{CR}$.}

First of all, we learn from Eq.~\eqref{eqn:vR_spiral} and Fig.~\ref{fig:spiral_grid} that a rigidly-rotating spiral perturbation 
wants to drive two signals ($\pm$) localized near the Lindblad resonances ($\omega^\mathrm{rel}_{m,\pm}\simeq 0$). The larger the lifetime $\tau$ of the spiral, the more narrowly focused each wave will be around its respective Lindblad resonance (see row (f) of Fig.~\ref{fig:spiral_grid}). On the other hand, the Gaussian in $(R-R_\mathrm{CR})$ penalizes each wave the further it is from corotation; this effect is stronger for small $\beta$, but largely unimportant for $\beta\gtrsim 0.5$ (row (e)). The role of pitch angle $\alpha$ is to set the curvature of the potential, which affects the overall wave amplitude (see row (d)). Indeed, assuming the $m\cot \alpha$ term dominates the final square bracket in \eqref{eqn:vR_spiral}, 
one can estimate the velocity perturbation near $R_\mathrm{g} \sim R_\mathrm{ref} \sim 8R_0$ to be $\vert \overline{v}_R / V_0\vert \lesssim \eta_\mathrm{eff}$, where
\begin{align}
    \eta_\mathrm{eff} &\equiv  2\pi \eta \,m \vert \cot \alpha \vert  \frac{\tau}{T_8}
    \label{eqn:eta_effective}
    \\
    &\simeq 0.05 \times \left( \frac{\eta}{0.005} \right ) \left( \frac{m}{2} \right ) \left( \frac{\cot \alpha}{2.5} \right )
    \left( \frac{\tau/T_8}{0.3} \right ).
        \label{eqn:vR_spiral_magnitude}
\end{align}
Next, at large $J_\varphi$, there is a rather simple phase offset between the signals arising from colder (blue) and hotter (red) populations. Physically this is just the effect of more rapid phase mixing, owing to the larger spread of (Dehnen-drifted) azimuthal frequencies for hotter populations (Eq.~\eqref{eqn:phase_shift}).
At smaller $J_\varphi$, there is not only a phase shift but also a suppression of the red compared to the blue signal. Mathematically, this arises from the imaginary argument of the {\tt erf} function in \eqref{eqn:vR_spiral}. Physically, it stems from resonance-detuning: the larger frequency spread in the hotter population means it responds less coherently to periodic forcing, and this is more apparent at low $J_\varphi$ where the dynamical clock is faster and the role of resonances more important.
Moving on to row (g), we see that for shearing spirals, the response is strongly concentrated around $R_\mathrm{CR}$.
Interestingly, at fixed $t$, increasing $\Gamma$ can lead to a larger amplitude \JphivR signal, because a more tightly-wound spiral tends to have a larger radial wavenumber $\vert k_R \vert$, strengthening the radial forcing.

\begin{figure}
    \centering
    \includegraphics[width=0.995\linewidth]{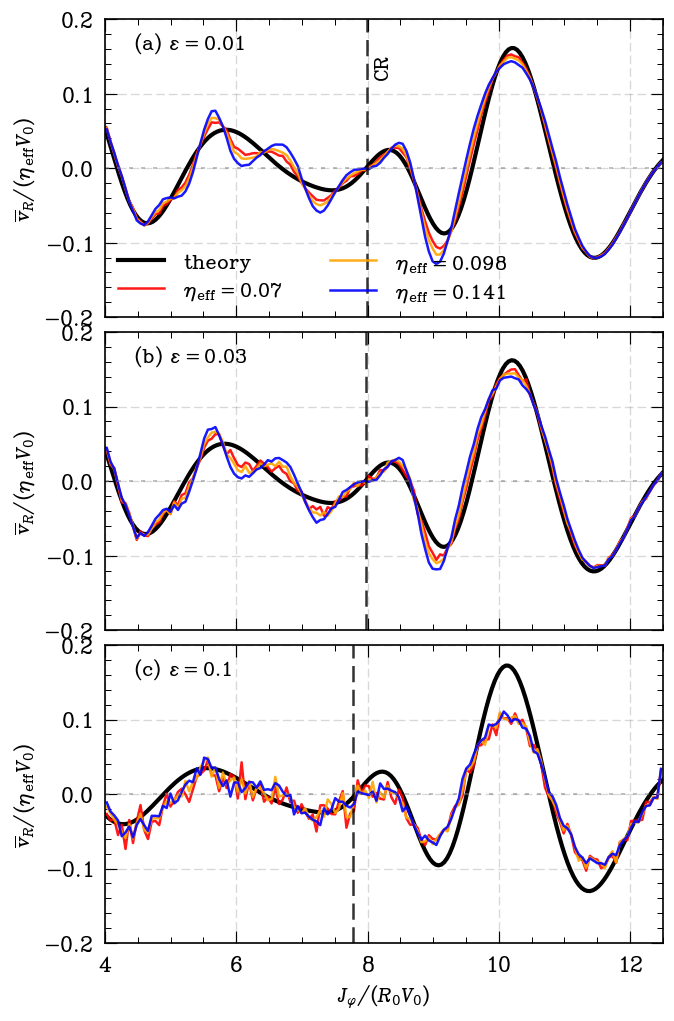}
    \caption{Similar to Figs.~\ref{fig:linearity_test} and \ref{fig:bar_lines}, except 
    using the transient spiral potential \eqref{eq:spiral_potential}, with $m=2$, $\alpha = 15^\circ$,  $R_\mathrm{CR}=R_\mathrm{ref}=8R_0$, $\beta=0.35$,
    $\varphi_0=0$, $\tau/T_8=0.3$ and $t/T_8=1.0$. Note that $\eta_\mathrm{eff} = 2\pi \eta m\vert \cot\alpha\vert \tau/T_8$ (see Eq. \eqref{eqn:vR_spiral_magnitude}).}
    \label{fig:spiral_lines}
\end{figure}

Finally, to test the validity of our linear theory solution \eqref{eqn:vR_spiral}
we present Fig.~\ref{fig:spiral_lines}. This is
analogous to Figs.~\ref{fig:linearity_test} and \ref{fig:bar_lines}, except we use the spiral perturbation \eqref{eq:spiral_potential} with the same parameters as in the fiducial (rigidly-rotating) model just described.
We also choose to normalize the signal by the value of $\eta_\mathrm{eff}$ (Eq.~\eqref{eqn:eta_effective}).
We see from Fig.~\ref{fig:spiral_lines} that at low $\eta$, our theoretical result is once again highly accurate for  $\epsilon \lesssim 0.03$, but it starts to break down for $\epsilon \gtrsim 0.1$ due to the long-wavelength assumption. 
By contrast, the theory begins to break down at the lowest $\epsilon$ if we make $\eta$ large enough, due to nonlinearity.
Importantly, though, nonlinear effects are \textit{less} important at higher $\epsilon$ (all the curves overlap in panel (c)).
This is because nonlinear effects are not only associated with a higher powers of $\eta$, but also with more derivatives of the background DF $f_0$ with respect to $J_R$, and the latter grows like $\propto \epsilon^{-2}$. In other words, at fixed $\eta$, a larger $\epsilon$ leads to a shallower gradient of $f_0$, and hence a smaller role is played by nonlinear terms.  This is yet more evidence that improving upon the long wavelength assumption, rather than the linear assumption, is what will render our theory more accurate.

\section{Fitting models to data}
\label{sec:fitting}

The key lesson we draw from the theory of \S\ref{sec:theory} and the examples of \S\ref{sec:Examples} is that, if the observed \JphivR signal can be understood as the result of a single perturbation to our Galactic disk, then a transient spiral is the most likely culprit. (The \textit{origin} of this spiral is a different question --- see the Discussion). 
It is therefore natural to ask whether a set of spiral parameters can be found whose resulting \JphivR signal matches the data in detail.  

The data we are interested in fitting was already shown with a black curve in Fig.~\ref{fig:best}. Following \cite{cao2024radial} (see the blue curve in their Fig.~1), this data was taken from Gaia DR3, focusing on an azimuthal wedge centered on the Sun $\vert \varphi - \varphi_\odot\vert < 0.2$,
calculating distances using \texttt{StarHorse} \citep{anders2022photo}, 
assuming \cite{bovy2015galpy}'s \texttt{MWPotential2014} as the background Galactic potential, and ignoring selection effects (and, in this case, keeping only cold orbits by imposing a vertical action cut $J_z < 3$ kpc km s$^{-1}$).
We emphasize that modifying any of these choices would produce a slightly different \JphivR curve, as can be seen by comparing \JphivR plots derived from the data by different groups (e.g. \citealt{friske2019more,cao2024radial,hunt2025milky,bernet2025dynamics}). Roughly speaking, as long as one makes sensible choices the peaks/troughs of the \JphivR signal are at essentially the same $J_\varphi$ locations but their amplitudes can be shifted by $\sim$ a few km s$^{-1}$ in $\overline{v}_R$. This shift is certainly not negligible, given that we are trying to fit fluctuations in $\overline{v}_R$.
An exploration of these issues is beyond the scope of this paper, and from now on we simply take the black curve in Fig.~\ref{fig:best}
as `the' \JphivR wave, but we must keep in mind that systematic errors can be significant.

We now perform a simple fit of our analytic model to the data.
We generate model \JphivR curves following the same procedure we described in \S\ref{sec:spiral}, for the azimuthal range $\vert \varphi \vert < 0.2$ (putting the Sun at $\varphi_\odot=0$ and setting $R_\mathrm{ref}=8$ kpc without loss of generality). We assume a rigidly-rotating spiral, so the analytic model is just Eq.~\eqref{eqn:vR_spiral} averaged over the azimuthal wedge. We fix the number of spiral arms to $m=2$.
For analytic ease we continue to assume a flat rotation curve with $V_0=220$ km s$^{-1}$, which is very close to, but not exactly the same as, \texttt{MWPotential2014} in the region of interest. 
For the DF of our stellar population we again assume stars are orbiting in the midplane\footnote{The cut on the data of $J_z<3$ kpc km s$^{-1}$ implies a typical vertical excursion $z_\mathrm{max} \lesssim 300$ pc, so we do not expect vertical motion to play an important role.} and set a constant $\langle J_R \rangle = \epsilon^2 \times 8$ kpc $\times \,V_0$ for some choice of $\epsilon$. We also add to the right-hand side of \eqref{eqn:vR_spiral} a possible overall offset $u$ (cf. Eq.~\eqref{eqn:friske}).\footnote{The need to include this offset likely stems, at least partly, from systematic uncertainties in the way the data is processed (see \S\ref{sec:Discussion}).  We found that if we forced $u=0$ we got a much worse fit.}

With this, our analytic model $\overline{v}_R(J_\varphi \vert \,\bm{p})$ depends on the following eight parameters:
\begin{equation}
    \bm{p} \equiv (\varphi_0, t, \eta, R_\mathrm{CR}, \alpha, \tau, \beta, u).
    \label{eqn:params}
\end{equation}
We use the Markov Chain Monte Carlo (MCMC) package \texttt{emcee} \citep{foreman2013emcee}
to perform a maximum likelihood fit to the black curve in Fig.~\ref{fig:best}, sampled every $20$ kpc km s$^{-1}$ in $J_\varphi$.
We adopt a global uncertainty in $\overline{v}_R$ of $2$ km s$^{-1}$ and assume a Gaussian likelihood function.
We set uniform priors on $\varphi_0/\mathrm{rad} \in (-\pi,\pi)$, $t/\mathrm{Gyr} \in (-0.01,1.0)$, $\eta\in(0.0005, 0.1)$, $R_\mathrm{CR}/\mathrm{kpc}\in(6.0,11.0)$, $\alpha/^\circ \in(5.0, 35.0)$, $\tau/\mathrm{Myr} \in (10, 350)$, $\beta\in(0.1,0.75)$,
$u/($km s$^{-1})$ $\in(-15,15)$. After some experimentation by hand, we settled on an initial guess $\bm{p}$ 
for which the \JphivR wave had roughly the right structure.  We then ran our MCMC chain, sampling the posterior probability distribution using 300 walkers for 30,000 steps.

    \begin{figure}
    \centering
    \includegraphics[width=0.99\linewidth]{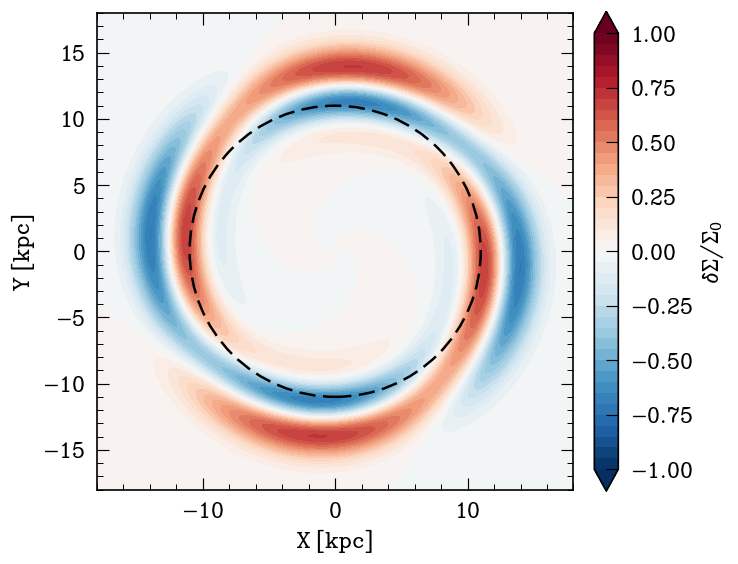}
    \caption{Dimensionless overdensity $\delta\Sigma/\Sigma_0$ corresponding to our best fit shearing spiral transient potential perturbation at $t=0$.  The black dashed line is $R_\mathrm{CR}=11.0$ kpc. We assume the spiral's vertical structure follows an isothermal profile with scale height $300$ pc, and that $\Sigma \propto \me^{-R/R_\mathrm{d}}$ with $R_\mathrm{d}=2.5$ kpc and $\Sigma_\odot = 50 M_\odot/$pc$^2$.}
    \label{fig:spiral_best}
\end{figure}

We begin by assuming $\epsilon=0.065$, which is somewhat smaller than the true $\epsilon$ measured today for the population under study ($\epsilon \sim 0.1$). 
With this choice the best fit posteriors are: 

The best fit posteriors are: 
$\varphi_0/\mathrm{rad} = -0.721_{-0.015}^{+0.015}$, 
$t/\mathrm{Gyr} = 0.3960_{-0.0003}^{+0.0003}$, 
$\eta = 0.075_{-0.002}^{+0.001}$, 
$R_\mathrm{CR}/\mathrm{kpc} = 11.000_{-0.004}^{+0.002}$, 
$\alpha/^\circ = 39.7_{-0.7}^{+0.8}$, 
$\tau/\mathrm{Myr} = 10.72_{-0.15}^{+0.32}$, 
$\beta = 0.203_{-0.002}^{+0.002}$, and 
$u/(\mathrm{km\,s}^{-1}) = 7.19_{-0.08}^{+0.08}$,
where we quote the posterior median and the corresponding $68\%$ confidence intervals.
We show the posterior distribution with a corner plot in Appendix \ref{sec:Corner} (red curves).  
The uncertainties here are much smaller than the assumed measurement error.  
Of course, we should keep in mind that these uncertainties are purely statistical, whereas the dominant source of uncertainty is almost certainly systematic, as we discuss in \S\ref{sec:Discussion}.
To check that this spiral could be generated by a sensible density perturbation, in Fig.~\ref{fig:spiral_best} we show 
the corresponding dimensionless overdensity $\delta\Sigma/\Sigma_0$ at the peak time ($t=0$), under the assumption that the spiral's 
vertical structure follows an isothermal profile with scale height $300$ pc, and that the background disk is exponential, $\Sigma \propto \me^{-R/R_\mathrm{d}}$ with $R_\mathrm{d}=2.5$ kpc and $\Sigma_\odot = 50$ M$_\odot$ pc$^{-2}$.  The typical over/underdensity here is a few tens of percent, so the spiral is at least physically realizable. 

These best fit parameters give rise to the red analytic theory curve shown in Fig.~\ref{fig:best}a.
This red curve agrees rather well with the data, matching the $J_\varphi$ locations of all the prominent peaks/troughs to within $\sim 50$ kpc km s$^{-1}$ and their $\overline{v}_R$ amplitudes to within a few km s$^{-1}$, as illustrated further in the residual plot, Fig.~\ref{fig:best}b. We confirmed the validity of the solution by performing an $N=5\times 10^7$ test particle simulation with these parameters, just as in \S\ref{sec:Examples}. The result is the blue line in Fig.~\ref{fig:best}, which shows that theory and simulation agree well. Quantitatively, both theory and simulation have smaller $\chi^2$ errors than the fitting formula \eqref{eqn:friske}, despite using two fewer free parameters.

However, we are concerned about the spiral lifetime $\tau$, which seems rather short. For instance, it is about a third of what one would expect for a swing amplification event centered on $R_\mathrm{CR}$\footnote{We deduced this by fitting a Gaussian function to Fig.~3 in
\cite{binney2020shearing}, which plots the amplitude of a surface density perturbation $\vert \delta \Sigma(t)\vert$ during swing amplification.
The standard deviation of this Gaussian is roughly $\tau \simeq 30$ Myr $\times\, R/(8 \, \mathrm{kpc})$}, and much shorter than expected for the growth and decay of a global instability.
We tried to find a solution with a more physically realistic $\tau$, but this tended to create \JphivR signals with wavelengths that were much too long. It is possible that by, e.g., increasing the number of spiral arms one may be able to find a fit that overcomes this issue (see row (a) of Fig.~\ref{fig:spiral_grid}), but our attempts failed. Instead, the fact that the data `prefers' low $\tau$ may really be hinting that our model is missing some key physics (see \S\ref{sec:Discussion}). 
We also note that the posterior distribution on $R_\mathrm{CR}$ seems to run into the upper bounds of the assumed prior. We tried allowing much higher maximum $R_\mathrm{CR}$ values, but the MCMC kept finding solutions whose corresponding $\delta \Sigma$ value was unphysical. We did not explore this further.

    \begin{figure}
    \centering
    \includegraphics[width=0.99\linewidth]{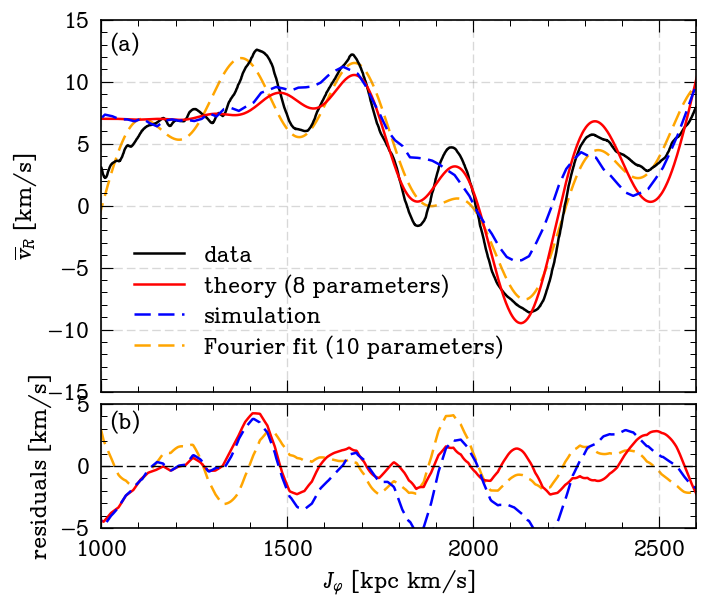}
    \caption{As in Fig.~\ref{fig:best} except for initial $\epsilon=0.1$. The theoretical curve agrees nicely with the data, but the solution is not valid since it is not well-matched by the test particle simulation result. This is because we have ignored important terms in the linear theory, which is only allowable for very small $\epsilon$.}
    \label{fig:spiral_pt1}
\end{figure}

Finally, we tried a run with $\epsilon=0.1$, which is a more appropriate value for the stellar population we are trying to model. The result is illustrated in Fig.~\ref{fig:spiral_pt1}. The best fit theory curve fits the data about as well as in
the $\epsilon=0.65$ case just discussed. The posterior distribution is also very similar (see Appendix \ref{sec:Corner}), with the only significant difference being an offset in the value of $\varphi_0$, as one might expect from the phase shift accrued by increasing $\langle J_R\rangle$ (e.g., Eq. \eqref{eqn:phase_shift}).
However, the corresponding test particle simulation for $\epsilon=0.1$ does not agree so well with the theory.
This disagreement may have been expected from Fig.~\ref{fig:spiral_lines}c --- it again stems from our use of the long wavelength assumption to (over-)simplify the mathematics. 

Thus, it seems likely that the best fit solutions we have found here are not exactly correct. 
They only fit the data accurately if one chooses an initial $\epsilon$ that is slightly too low and the spiral lifetime $\tau$ is unrealistically short.
Nevertheless, the methodology we have developed is clearly capturing the bulk of the correct physics and is
capable of exploring parameter space efficiently.

\section{Discussion}
\label{sec:Discussion}

The \JphivR wave observed in the Galactic disk is a relatively weak fluctuation ($\vert v_R /V_0\vert \sim 5\%$) that probably arose $\lesssim 1$ Gyr ago, and as such is a prime target for perturbation theory techniques. Taking this point of view, we have developed an analytic theory for the \JphivR wave, culminating in Eq.~\eqref{eqn:meanvr2}.  This theory is based on the following two key approximations. (i) The potential fluctuation that caused the \JphivR wave must be sufficiently weak (small enough $\eta$) 
that its behavior can be described with linear theory. 
(ii) The population of stars under investigation must be sufficiently dynamically cold (small enough $\epsilon$) that those potential perturbations are in the long wavelength regime, allowing us to simplify the linear theory dramatically.
The numerical tests we performed in \S\ref{sec:Examples}
show that assumption (ii) is the most restrictive. More precisely, our theory is very 
accurate for very cold populations with $\epsilon \ll 0.1$, but can break down for $\epsilon \gtrsim 0.1$ (radial velocity dispersion $\sigma \gtrsim 30$ km s$^{-1}$ --- see Eq.~\eqref{eqn:sigma}).
On the other hand, even for $\epsilon \gtrsim 0.1$ the theory still tends to provide good qualitative insight (e.g., Fig.~\ref{fig:spiral_lines}).

In practice we also made another important assumption, namely that (iii) we may ignore the self-gravity of the perturbed stellar distribution in our calculations. In other words, we did not impose any Poisson equation. 
One should rightly worry that this throws the validity of the entire enterprise into doubt.
On the other hand, self gravity \textit{is} implicitly included in the calculation, \textit{if} we guess the full potential perturbation (including self-gravity) correctly.
To be more concrete, let us imagine the following experiment.
We run a self-consistent N-body simulation, and extract the full potential fluctuation, which we call  $\delta \phi_1(\bm{r}, t)$.  Then, we rerun the same simulation {applying} $\delta \phi_1(\bm{r}, t)$ as an external forcing, and we
extract the  full potential fluctuation from this simulation, call it $\delta \phi_2(\bm{r}, t)$. If $\delta \phi_2(\bm{r}, t) \propto \delta \phi_1(\bm{r}, t)$, then our analytic theory is valid ---
any such $\delta \phi_1(\bm{r}, t)$ can be inserted into Eq.~\eqref{eqn:meanvr2}  without worrying that we are neglecting self-gravity, because we have guessed the full potential fluctuation correctly (up to an overall proportionality constant).
Mathematically, the condition we are imposing is roughly that the potential response be dominated by Landau modes \citep{hamilton2024kinetic}. The linear response of isolated stellar disks in $N$-body simulations can be dominated by these  modes \citep{sellwood2014transient,hamilton2024kinetic,roule2025long}, so this may not be a bad assumption.  However, whether this approach is valid in realistic galactic contexts is an open question.
It is likely invalid in the case of perturbations by Sgr, where one should really compute the bodily acceleration of the entire disk in response to Sgr's infall \citep{binney2024disc}, rather than just the in-plane fluctuations.

We also ignored the role of the gaseous interstellar medium (ISM). Fluctuations in gravitational potential due to the ISM can scatter stellar orbits, and depending on the circumstances this can either preserve \citep{chiba2025origin} or destroy \citep{tremaine2023origin,tepper2025galactic} phase space substructures. We expect that simple models of this scattering could be built into the analytic formalism \citep{hamilton2023galactic,modak2025characterizing}, but as of yet we can make no statement on how the ISM affects the \JphivR wave.

Setting these issues aside, both theoretical and numerical evidence in \S\ref{sec:Examples} suggested that transient spiral structure is the most likely cause of the \JphivR wave, and ultimately led to the best-fit solution plotted in Fig.~\ref{fig:best} and discussed in \S\ref{sec:fitting}.  The fit of analytic theory to data is very good, far better than we have seen in other models, and we are confident that the bulk of the physics is being captured correctly.
Nevertheless, there are several limitations and caveats to this story that should be borne in mind before drawing 
any firm conclusions.

To begin with, one should be skeptical about the fit found in Fig.~\ref{fig:best}. 
We are unable to say why the corresponding spiral transient would have arisen, and 
the best fit lifetime $\tau$ seems problematically short.  On the other hand, 
we experimented with simple barred models, and with higher-$m$ spirals, but they all fared much worse.  Even a theory-agnostic, \textit{ten}-parameter fit of three sine waves and an offset --- see Eq.~\eqref{eqn:friske} and the gold dashed line in Fig.~\ref{fig:best} ---  does not do a better job than our eight-parameter spiral.  This suggests that our model, far from being infinitely flexible, is actually close to an optimal compression of the information required to reproduce the observed signal within a tolerable error. So, while there are many issues remaining, we do believe we are on the right track.

More importantly, our theory and simulations only agree nicely if we pick an initial condition that is slightly too cold compared to the populations we are interested in. More precisely, the theory breaks down quantitatively when $\epsilon\gtrsim 0.1$ (although it is still qualitatively valuable).
Thus, a theoretical model following \eqref{eqn:meanvr2} that fits the data nicely for these $\epsilon$ values may be misleading, since it may not be reproducible in reality (see, e.g., Fig.~\ref{fig:spiral_pt1}). 
On the other hand, it will be straightforward to extend the theory to include next-order corrections to \eqref{eqn:meanvr2}, and we suspect this extended theory will be accurate over a much more realistic $\epsilon$ range. We leave this for future work.

Another important concern is systematic error induced by the choice of background potential, local standard of rest, how to calculate distances, etc.
Making different choices leads to a slight distortion of both the `observed' \JphivR curve \textit{and} the theoretical prediction. One must also account for selection biases in the data that predominantly affect hotter orbits (e.g., \citealt{bland2019}), but we ignored this issue entirely here. These choices are not arbitrary, because the relative amplitudes of the fluctuations we are trying to understand can sometimes be comparable to the systematic uncertainty inherent in them. Thus, one should not take any fit found by our method too seriously until these issues are treated carefully.

Lastly, there is a wealth of kinematic and chemical data beyond the simple one-dimensional \JphivR wave that we have concentrated on, including wrinkles and corrugations in the out-of-plane variables \citep{schonrich2018warp,bland2021galactic,hunt2025milky}. In principle a complete model of our Galaxy's equilibrium DF \citep{binney2023self,binney2024chemodynamical} and recent history of perturbations should account for all of it. We have not checked the extent to which our transient spiral solution is consistent with any of these data. Doing so is a clear avenue for future work.

\section{Summary}
\label{sec:Summary}

In this paper, we have studied the physical mechanism behind the \JphivR wave seen in the Galactic disk.
Our findings can be summarized as follows:
\begin{itemize}
    \item We used linear theory (\S\ref{sec:theory} and Appendix \ref{sec:derivation}) to derive an analytic formula  for the $J_\varphi$-$\overline{v}_R$ signal generated by a given potential perturbation $\delta \phi$ (Eq.~\eqref{eqn:meanvr2}), and tested it numerically for several examples (\S\ref{sec:Examples}).  The theory works excellently for dynamically cold stellar populations. For hotter populations it provides valuable qualitative insight but cannot always be trusted quantitatively.
    
    \item We found that any short-lived, weak, $m$-fold perturbation tends to excite two waves in \JphivR space, that rotate in azimuth $\varphi$ with effective frequencies $m \Omega \pm \kappa + 2m\kappa'\langle J_R \rangle $ (Eq.~\eqref{eqn:omegaeff}).
    Longer-lived perturbations also drive \JphivR waves but their frequency structure is more complicated, especially near resonances.
    
    \item In all cases, at a fixed $\varphi$, the `wavelength' in $J_\varphi$ space tends to increase with $J_\varphi$ and decrease with time (e.g., Eq.~\eqref{eqn:wavelength_Jphi}). Dynamically hotter (larger $\langle J_R \rangle$) stars exhibit a \JphivR wave that generically lags that of cold (smaller $\langle J_R \rangle$) populations, because more eccentric orbits have lower azimuthal frequencies (e.g., Eq.~\eqref{eqn:phase_shift}).

    \item Generically, different perturbations give rise to different \JphivR signals. In the absence of self-gravity, distant tidal kicks like that due to the Sgr dwarf tend to produce a signal dominated by a single sinusoidal component; forcing by a rigidly-rotating bar drives a signal dominated by two components; and transient spiral structure is capable of driving three-component waves, similar to what is observed (Eq.~\eqref{eqn:friske} and Fig.~\ref{fig:best}).
       
    \item We used a Markov Chain Monte-Carlo method to fit our analytic model to the data, assuming a transient spiral perturbation (\S\ref{sec:fitting}).  The best fit spiral drives a \JphivR wave quite consistent with observations (Fig.~\ref{fig:best}). However, various caveats and systematic effects need to be investigated carefully before the solution can be trusted  (\S\ref{sec:Discussion}).
\vspace{0.5em}
\end{itemize}

Perhaps the most important conclusion to be drawn from our work
is simply that the \JphivR problem is a \textit{linear} one, with nonlinearities typically playing a
sub-dominant role.
This suggests that a complete solution may well be found in the realm of linear perturbation theory. But if this is the case, the final linear theory will have to be more elaborate than ours, transcending at least one (and possibly all) of our key assumptions (i)-(iii).  Analytically tractable extensions to our theory are possible \citep{hamilton2026galactokinetics,GK2}, but ultimately one may wish to do the perturbation theory numerically.
Fortunately, there already exists significant machinery tailored to exactly this kind of work
\citep{binney2016torus,Binney2018orbital,binney2020angle,petersen2024predicting}. These authors' numerical methods are highly accurate and rely on very few approximations. 
Meanwhile, our analytic theory is extremely fast to execute, allowing rapid exploration of the parameter space.
Taken together, these complementary approaches may allow us to solve the \JphivR mystery --- and some of the other dynamical puzzles Gaia has bequeathed us --- once and for all.

\begin{acknowledgments}
C.H. and A.M. are supported by the John N. Bahcall Fellowship Fund at the Institute for Advanced Study. A.M. acknowledges support from the Ambrose Monell Foundation and the W.M. Keck Foundation.
C.H. thanks Chengye Cao and Marcel Bernet for alerting him to this problem.
\end{acknowledgments}


%


\appendix

\section{Derivation of the \JphivR formula}
\label{sec:derivation}

We follow closely the notation of \cite{hamilton2026galactokinetics}. 
In the (nearly-)epicyclic mapping used there, the radial velocity and azimuthal angle can be expressed as
\begin{equation}
    v_R = a_R \kappa \sin\theta_R, \,\,\,\,\,\,\,\,\,\,\,\,\,\,\,\,\,\, \mathrm{and} \,\,\,\,\,\,\,\,\,\,\,\,\,\,\,\,\,\,
    \varphi = \theta_\varphi + \gamma
     \frac{a_R}{R_\mathrm{g}}\sin\theta_R{+ \frac{1}{4}a_R^2\frac{\md \kappa}{\md J_\varphi}
     \sin 2\theta_R},
     \label{eqn:radvel}
\end{equation}
respectively.
The final term on the right-hand side accounts for the Dehnen drift.
Writing the distribution function (DF) of the full disk as $f(\btheta,\bJ,t)$, the number density of stars at a fixed $\varphi$ and $J_\varphi$ is then proportional to 
\begin{equation}
    N(\varphi, J_\varphi,t) \equiv \int \md \theta_\varphi \, \md J_\varphi' \, \md \theta_R \,\md J_R \, f(\theta_\varphi, \theta_R, J_\varphi', J_R)\,\delta(J_\varphi'-J_\varphi)\,\delta \left( \varphi - \bigg[\theta_\varphi + \gamma
     \frac{a_R}{R_\mathrm{g}}\sin\theta_R{+ \frac{1}{4}a_R^2\frac{\md \kappa}{\md J_\varphi}
     \sin 2\theta_R}
    \bigg]\right),
    \label{eqn:number_density}
\end{equation}
The mean radial velocity at $\varphi$, $J_\varphi$ is defined similarly, with a crucial extra factor of $v_R$ in the integrand:
\begin{align}
    \overline{v}_R(\varphi, J_\varphi,t) \equiv \frac{1}{N(\varphi, J_\varphi)} \int \md \theta_\varphi \, \md J_\varphi' \, \md & \theta_R \,\md J_R \, f(\theta_\varphi, \theta_R, J_\varphi', J_R) \nn
    \\
&    \times \,\delta(J_\varphi'-J_\varphi)\,\delta \left( \varphi - \bigg[\theta_\varphi + \gamma
     \frac{a_R}{R_\mathrm{g}}\sin\theta_R{+ \frac{1}{4}a_R^2\frac{\md \kappa}{\md J_\varphi}
     \sin 2\theta_R}
    \bigg]\right) a_R\kappa \sin\theta_R .
    \label{eqn:vRdefinition}
\end{align}

As in \cite{hamilton2026galactokinetics}, we now write $f(\btheta,\bJ,t) = f_0(\bJ, t) + \delta f(\btheta, \bJ, t)$, and decompose $\delta f$ as a Fourier series:
\begin{align}
    \label{eq:Fourier_deltaf}
    \delta f(\btheta, \bm{J}) = & \sum_{{\bm{n}}} \delta f_{{\bm{n}}}(\bm{J}) \, \me^{i {\bm{n}}\cdot\btheta},
\end{align}
where $\bn=(n_\varphi, n_R)$ are vectors of integers. We plug this decomposition into Eqs.~\eqref{eqn:number_density}-\eqref{eqn:vRdefinition}, arriving at 
\begin{align}
&   N(\varphi, J_\varphi,t) =  F_0(J_\varphi) + 2 \pi \sum_{\bn} \int \md J_R      \,\me^{in_\varphi\varphi } \, \delta f_{\bn}(\bJ, t)
     J_{n_R}\left( \frac{n_\varphi\gamma a_R}{R_\mathrm{g}}\right) + ...,
    \label{eqn:Nagain}
\end{align}
\begin{align}
&   \overline{v}_R(\varphi, J_\varphi,t) = \frac{1}{N(\varphi, J_\varphi)} 2 \pi \sum_{\bn} \int \md J_R  \frac{a_R\kappa}{2i}    \,\me^{in_\varphi\varphi } \, \delta f_{\bn}(\bJ, t)
    \bigg[ J_{n_R+1}\left( \frac{n_\varphi\gamma a_R}{R_\mathrm{g}}\right)
    - 
    J_{n_R-1}\left( \frac{n_\varphi\gamma a_R}{R_\mathrm{g}}\right) + ...
    \bigg],
    \label{eqn:meanvr2again}
\end{align}
where $F_0(J_\varphi) \equiv 2\pi \int_0^\infty \md J_R \,f_0(\bJ)$.
The ellipsis in these equations corresponds to terms arising from the $\sin 2\theta_R$ terms in the delta functions in \eqref{eqn:number_density}-\eqref{eqn:vRdefinition}. We will not write them out here because they turn out to be negligible --- of order $\mathcal{O}(\epsilon_R^2)$ compared to the terms we will keep --- for reasons that will become clear momentarily.

The majority of perturbations we are interested in have all their power invested in small azimuthal wavenumbers $\vert n_\varphi\vert \sim 1$. Given this, the arguments of the relevant Bessel functions in \eqref{eqn:Nagain}-\eqref{eqn:meanvr2again} are $\mathcal{O}(\epsilon_R)$, so we can expand them using $J_\ell(x) \simeq (x/2)^\ell/\ell!$:
\begin{align}
&  N(\varphi, J_\varphi,t) \simeq F_0 + 2\pi \sum_{n_\varphi} \int \md J_R    \,\me^{in_\varphi\varphi } \,
   \bigg[\delta f_{n_\varphi, 0} +  \frac{n_\varphi \gamma a_R}{2 R_\mathrm{g}} \left(  \delta f_{n_\varphi,1} - \delta f_{n_\varphi,-1} \right) + ... \bigg] ,
    \label{eqn:Npre}
\end{align}
\begin{align}
&    \overline{v}_R(\varphi, J_\varphi,t) \simeq \frac{1}{N(\varphi, J_\varphi)} 2 \pi \sum_{n_\varphi} \int \md J_R  \frac{a_R\kappa}{2i}    \,\me^{in_\varphi\varphi } \,
   \bigg[\delta f_{n_\varphi, 1} - \delta f_{n_\varphi,-1} +  \frac{\gamma a_R}{R_\mathrm{g}} \left( \delta f_{n_\varphi,0} + \delta f_{n_\varphi,2} - \delta f_{n_\varphi,2} \right) + ...\bigg].
    \label{eqn:meanvr2pre}
\end{align}
Interestingly, \eqref{eqn:meanvr2pre} tells us that the mean radial velocity
$\overline{v}_R$ has contributions from components of the 
DF that are \textit{even} in radial angle ($n_R=0,\pm 2$), despite the fact that $v_R$ itself is \textit{odd} in radial angle (Eq.~\eqref{eqn:radvel}).  This stems from the fact that  we have chosen to look at a {fixed} $\varphi$, as is most appropriate for comparing to observations.  Had we instead worked at a fixed $\theta_\varphi$, or looked at the azimuthally-averaged signal, 
these even terms would not arise. 

Equations \eqref{eqn:Npre}-\eqref{eqn:meanvr2pre} tell us that
to calculate $\overline{v}_R$, we need to know the Fourier components $\delta f_{n_\varphi,n_R}(\bJ, t)$ for $n_R=0,\pm 1, \pm2$.
We will calculate these components using linear response theory: we treat $f_0(\bJ)$ as fixed, and consider the evolution of the (presumed small) fluctuation $\delta f$ in the presence of a weak potential perturbations $\delta \phi$, which may or may not include a self-consistent piece.  
  The central result of linear theory is (e.g., \citealt{hamilton2026galactokinetics}):
 \begin{align}
    \label{eq:linear_Vlasov_formal_solution}
    &\delta f_{{\bm{n}}}(\bm{J}, t) =  \delta f_{{\bm{n}}}(\bm{J}, 0) \me^{-i\bn\cdot\bOm t}  +  i {\bm{n}}\cdot \frac{\p f_0}{\p \bm{J}}\int_0^t \md t' \, \me^{- i {\bm{n}}\cdot \bOm (t-t')} \delta\phi_{{\bm{n}}}(\bm{J}, t').
\end{align}
The first term on the right-hand side 
encodes pure phase mixing of any angle-dependent initial fluctuation in the DF, while the second term tells us how additional DF fluctuations are generated by potential perturbations.

We can make some further simplifications as follows. First, we assume the disk is initially unperturbed, $\delta f_{\bn}(\bJ,0)=0$. Of course, this is not generally the case, but it must be approximately true if we are to have any hope of pinning down $\delta \phi$ given the data on $\overline{v}_R$. 
Secondly, we use the fact that the DF $f_0$ is usually a much steeper function of $J_R$ than it is of $J_\varphi$; typically
$\vert \p f_0/\p J_\varphi\vert  \sim  \epsilon_R^2 \vert \p f_0 /\p J_R \vert $. 
Thirdly, the perturbations we care about in this context --- bars, spirals, and distant dwarf galaxy encounters --- are typically of long or intermediate wavelength \citep{hamilton2026galactokinetics}, so we can approximate the Fourier components $\delta \phi_{\bn}$ using Eq.~(74) of \cite{hamilton2026galactokinetics}. In particular this means that every potential fluctuation component $\delta \phi_{\bn}$ with non-zero $n_R$ is suppressed by $\sim \epsilon_R^{\vert n_R\vert}$ relative to the $n_R=0$ component. 

Substituting \eqref{eq:linear_Vlasov_formal_solution} into \eqref{eqn:Npre} and using these approximations, it turns out that the terms involving $\delta f$ terms are subdominant (i.e., $\mathcal{O}(\epsilon_R^2)$) compared to $F_0$, so we just have $N\simeq F_0$.

Similarly, plugging \eqref{eq:linear_Vlasov_formal_solution} into \eqref{eqn:meanvr2pre}, it turns out that the $n_R=\pm 1$ terms are dominant (the $n_R = 0$ terms are smaller by $\mathcal{O}(\epsilon_R)$, while the $n_R=\pm 2$ terms are smaller by $\mathcal{O}(\epsilon_R^2)$). The `...' terms that we dropped in Eqs.~\eqref{eqn:Nagain}-\eqref{eqn:meanvr2again} are all negligible for the same reasons. 
We drop all such subdominant terms from now on, although they could easily be included in a more accurate analysis. We also assume  $f_0$ has the Schwarzschild-type form \eqref{eqn:Schwarzchild}, allowing us to
perform $J_R$ integrals explicitly. Since $N\simeq F_0$, the $F_0$ dependence cancels out in equation \eqref{eqn:meanvr2pre}.  Using $\delta f_{-\bn} = \delta f_{\bn}^*$ and  $\delta \Phi_{-m}=\delta\Phi^*_m$, 
we soon arrive at an explicit formula for the \JphivR wave:
\begin{align}
    \overline{v}_R(\varphi, J_\varphi, t) \simeq \mathrm{Re} \sum_{m=1}^\infty 
    \sum_{\pm} \,\me^{i[m\varphi-(m\Omega\pm\kappa) t]} \int_0^t \md t' \frac{ \me^{i (m\Omega\pm\kappa) t'}}{[1+im \kappa' \langle J_R \rangle (t-t')]^2}  \delta\Phi_{m}(t')  \left[\pm \frac{\gamma m}{R_\mathrm{g}} - \mi k_R(t')\right].
   \label{eqn:meanvr2yet}
\end{align}

One final simplification is possible, which is not strictly necessary but is convenient for performing explicit analytic calculations (see, e.g., \S\ref{sec:Examples}). Namely, the term involving $\langle J_R\rangle$ in the denominator of the integrand arises because of the Dehnen drift term $\Omega_\mathrm{D}$ in the azimuthal frequency, and 
this term is often much smaller than unity, so we can write 
\begin{equation}
   [1+im \kappa' \langle J_R \rangle (t-t')]^{-2} \simeq \me^{-2im \kappa' \langle J_R \rangle (t-t')}.
\label{eqn:Dehneneasier}
\end{equation}
  Corrections to the approximation \eqref{eqn:Dehneneasier} will be $\mathcal{O}(x^2)$
where
\begin{equation}
x \equiv \vert m \kappa' \langle J_R \rangle (t-t') \vert \sim m \epsilon^2 \Omega (t-t') \sim 0.1 \times \left( \frac{m}{2} \right) \left( \frac{\epsilon}{0.1} \right)^2 \left( \frac{t-t'}{200 \, \mathrm{Myr}} \right) \left( \frac{T_\varphi}{200 \, \mathrm{Myr}} \right)^{-1}.
\end{equation}
Thus, \eqref{eqn:Dehneneasier} is accurate to the level of a few percent, provided we restrict our analysis to timescales smaller than $\sim 1$ Gyr.
Using \eqref{eqn:Dehneneasier} in \eqref{eqn:meanvr2yet}, we arrive at the final result \eqref{eqn:meanvr2}.
Note that to this order, the role of the Dehnen drift is just to drive a shift $2m\kappa'\langle J_R \rangle$ in the effective frequency of the response (equation \eqref{eqn:omegaeff}).

\section{MCMC details}
\label{sec:Corner}

\begin{figure}
    \centering
    \includegraphics[width=0.995\linewidth]{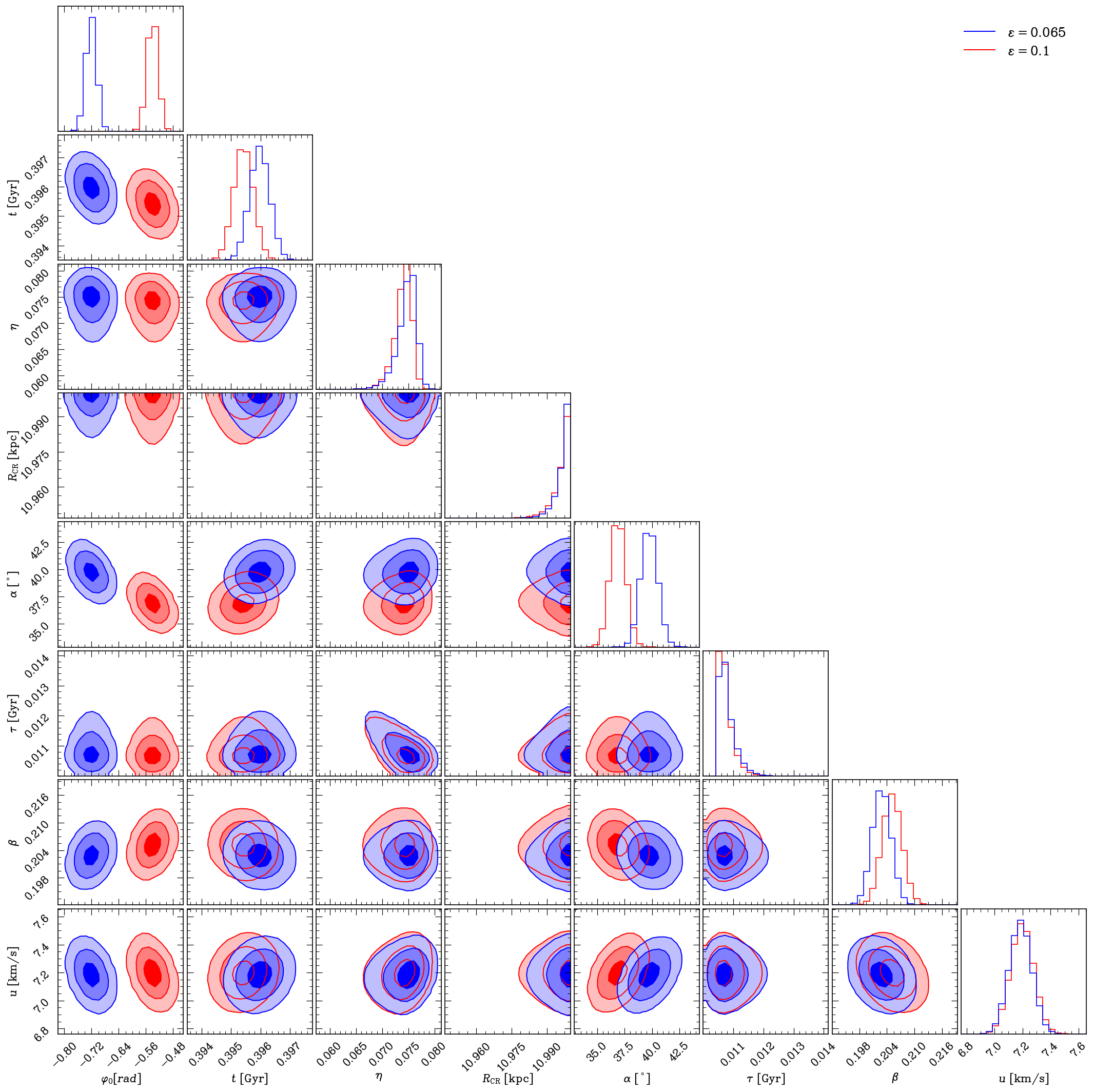}
    \caption{Corner plot resulting from the two MCMC fits discussed in \S\ref{sec:fitting}. 
    The blue (red) curves correspond to the model with $\epsilon = 0.065$ ($\epsilon=0.1$).}
    \label{fig:Corner}
\end{figure}

In Figure \ref{fig:Corner} we show the corner plot of the posterior distribution resulting from the MCMC fits discussed in \S\ref{sec:fitting}.

\bibliography{sample}{}

@article{binney2023self,
  title={Self-consistent models of our Galaxy},
  author={Binney, James and Vasiliev, Eugene},
  journal={MNRAS},
  volume={520},
  number={2},
  pages={1832--1847},
  year={2023},
  publisher={Oxford University Press}
}

@ARTICLE{Monari2019-er,
  title     = "Signatures of the resonances of a large Galactic bar in local velocity space",
  author = {{Monari}, G. and {Famaey}, B. and {Siebert}, A. and {Wegg}, C. and {Gerhard}, O.},
  journal   = "A\&A",
  publisher = "aanda.org",
        year = 2019,
        month = jun,
       volume = {626},
          eid = {A41},
        pages = {A41}
}

@article{khalil2025non,
  title={A non-axisymmetric potential for the Milky Way disk},
  author={Khalil, YR and Famaey, B and Monari, G and Bernet, M and Siebert, A and Ibata, R and Thomas, GF and Ramos, P and Antoja, T and Li, C and others},
  journal={Astronomy \& Astrophysics},
  volume={699},
  pages={A263},
  year={2025},
  publisher={EDP Sciences}
}

@article{binney2024disc,
  title={Disc distortion revisited},
  author={Binney, James},
  journal={MNRAS},
  volume={535},
  number={2},
  pages={1898--1912},
  year={2024},
  publisher={Oxford University Press}
}

@article{petersen2024predicting,
  title={Predicting the linear response of self-gravitating stellar spheres and discs with {L}inearResponse.jl},
  author={Petersen, Michael S and Roule, Mathieu and Fouvry, Jean-Baptiste and Pichon, Christophe and Tep, Kerwann},
  journal={\mnras},
  volume={530},
  number={4},
  pages={4378--4394},
  year={2024},
  publisher={Oxford University Press}
}

@article{binney2020angle,
  title={Angle-action variables for orbits trapped at a Lindblad resonance},
  author={Binney, James},
  journal={\mnras},
  volume={495},
  number={1},
  pages={886--894},
  year={2020},
  publisher={Oxford University Press}
}

@article{Binney2018orbital,
  title={Orbital tori for non-axisymmetric galaxies},
  author={Binney, James},
  journal={\mnras},
  volume={474},
  number={2},
  pages={2706--2724},
  year={2018},
  publisher={Oxford University Press}
}

@article{dehnen1999approximating,
  title={Approximating Stellar Orbits: Improving on EpicycleTheory},
  author={Dehnen, Walter},
  journal={AJ},
  volume={118},
  number={3},
  pages={1190},
  year={1999},
  publisher={IOP Publishing}
}

@article{binney2016torus,
  title={Torus mapper: a code for dynamical models of galaxies},
  author={Binney, James and McMillan, Paul J},
  journal={MNRAS},
  volume={456},
  number={2},
  pages={1982--1998},
  year={2016},
  publisher={The Royal Astronomical Society}
}

@article{roule2025long,
  title={The long-term evolution of razor-thin galactic discs: Balescu--Lenard prediction and perspectives},
  author={Roule, Mathieu and Fouvry, Jean-Baptiste and Pichon, Christophe and Chavanis, Pierre-Henri},
  journal={A\&A},
  volume={699},
  pages={A140},
  year={2025},
  publisher={EDP Sciences}
}

@article{julian1966non,
  title={Non-axisymmetric responses of differentially rotating disks of stars},
  author={Julian, William H and Toomre, Alar},
  journal={ApJ},
  volume={146},
  pages={810},
  year={1966}
}

@article{anders2022photo,
  title={Photo-astrometric distances, extinctions, and astrophysical parameters for Gaia EDR3 stars brighter than G= 18.5},
  author={Anders, Friedrich and Khalatyan, A and Queiroz, Anna B{\'a}rbara de Andrade and Chiappini, Cristina and Ard{\`e}vol, J and Casamiquela, Laia and Figueras, Francesc and Jim{\'e}nez-Arranz, {\'O} and Jordi, Carme and Mongui{\'o}, M and others},
  journal={A\&A},
  volume={658},
  pages={A91},
  year={2022},
  publisher={EDP Sciences}
}

@article{tepper2025galactic,
  title={Galactic seismology: can the Gaia ‘phase spiral’co-exist with a clumpy, turbulent interstellar medium?},
  author={Tepper-Garc{\'\i}a, Thor and Bland-Hawthorn, Joss and Bedding, Timothy R and Federrath, Christoph and Agertz, Oscar},
  journal={MNRAS},
  volume={542},
  number={3},
  pages={1987--2003},
  year={2025},
  publisher={Oxford University Press}
}

@article{tremaine2023origin,
  title={The origin and fate of the Gaia phase-space snail},
  author={Tremaine, Scott and Frankel, Neige and Bovy, Jo},
  journal={MNRAS},
  volume={521},
  number={1},
  pages={114--123},
  year={2023},
  publisher={Oxford University Press}
}

@article{schonrich2018warp,
  title={Warp, waves, and wrinkles in the Milky Way},
  author={Sch{\"o}nrich, Ralph and Dehnen, Walter},
  journal={MNRAS},
  volume={478},
  number={3},
  pages={3809--3824},
  year={2018},
  publisher={Oxford University Press}
}

@article{chiba2021resonance,
  title={Resonance sweeping by a decelerating Galactic bar},
  author={Chiba, Rimpei and Friske, Jennifer KS and Sch{\"o}nrich, Ralph},
  journal={\mnras},
  volume={500},
  number={4},
  pages={4710--4729},
  year={2021},
  publisher={Oxford University Press}
}

@ARTICLE{Chiba2022-qt,
  title     = "Oscillating dynamical friction on galactic bars by trapped dark
               matter",
  author    = "Chiba, Rimpei and Sch{\"o}nrich, Ralph",
  journal   = "\mnras",
  publisher = "Oxford Academic",
  volume    =  513,
  number    =  1,
  pages     = "768--787",
  month     =  mar,
  year      =  2022,
  language  = "en"
}

@article{sellwood2014transient,
  title={Transient spirals as superposed instabilities},
  author={Sellwood, JA and Carlberg, RG},
  journal={\apj},
  volume={785},
  number={2},
  pages={137},
  year={2014},
  publisher={IOP Publishing}
}

@article{hamilton2023galactic,
  title={Galactic Bar Resonances with Diffusion: An Analytic Model with Implications for Bar--Dark Matter Halo Dynamical Friction},
  author={Hamilton, Chris and Tolman, Elizabeth A and Arzamasskiy, Lev and Duarte, Vin{\'\i}cius N},
  journal={\apj},
  volume={954},
  number={1},
  pages={12},
  year={2023},
  publisher={IOP Publishing}
}

@article{bovy2015galpy,
  title={galpy: A python Library for Galactic Dynamics},
  author={Bovy, Jo},
  journal={ApJ Supplement Series},
  volume={216},
  number={2},
  pages={29},
  year={2015},
  publisher={IOP Publishing}
}

@ARTICLE{GK2,
       author = {{Hamilton}, Chris and {Modak}, Shaunak and {Tremaine}, Scott},
        title = "{Galactokinetics II: Spiral structure}",
      journal = {arXiv e-prints},
         year = 2025,
        month = jul,
        pages = {arXiv:2507.16950},
       eprint = {2507.16950},
 primaryClass = {astro-ph.GA}
}

@ARTICLE{Grivnev1988,
       author = {{Grivnev}, E.~M.},
        title = "{Galactic Star Orbits in the POST Epicyclic Approximation}",
      journal = {\sovast},
         year = 1988,
        month = apr,
       volume = {32},
        pages = {139},
       adsurl = {https://ui.adsabs.harvard.edu/abs/1988SvA....32..139G}
}

@article{binney2020shearing,
  title={The shearing sheet and swing amplification revisited},
  author={Binney, James},
  journal={MNRAS},
  volume={496},
  number={1},
  pages={767--783},
  year={2020},
  publisher={Oxford University Press}
}

@article{antoja2022tidally,
  title={Tidally induced spiral arm wraps encoded in phase space},
  author={Antoja, Teresa and Ramos, Pau and L{\'o}pez-Guitart, Ferran and Anders, Friedrich and Bernet, Marcel and Laporte, Chervin FP},
  journal={A\&A},
  volume={668},
  pages={A61},
  year={2022},
  publisher={EDP Sciences}
}

@article{hunt2020power,
  title={The power of coordinate transformations in dynamical interpretations of Galactic structure},
  author={Hunt, Jason AS and Johnston, Kathryn V and Pettitt, Alex R and Cunningham, Emily C and Kawata, Daisuke and Hogg, David W},
  journal={MNRAS},
  volume={497},
  number={1},
  pages={818--828},
  year={2020},
  publisher={Oxford University Press}
}

@article{dehnen2000effect,
  title={The Effect of the Outer Lindblad Resonance of the Galactic Baron the Local Stellar Velocity Distribution},
  author={Dehnen, Walter},
  journal={AJ},
  volume={119},
  number={2},
  pages={800},
  year={2000},
  publisher={IOP Publishing}
}

@article{bernet2025dynamics,
  title={Dynamics of tidal spiral arms: Machine learning-assisted identification of equations and application to the Milky Way},
  author={Bernet, Marcel and Ramos, Pau and Antoja, Teresa and Price-Whelan, Adrian and Brunton, Steven L and Asano, Tetsuro and Gir{\'o}n-Soto, Alexandra},
  journal={arXiv preprint arXiv:2506.17383},
  year={2025}
}

@article{cao2024radial,
  title={Radial Wave in the Galactic Disk: New Clues to Discriminate Different Perturbations},
  author={Cao, Chengye and Li, Zhao-Yu and Sch{\"o}nrich, Ralph and Antoja, Teresa},
  journal={ApJ},
  volume={975},
  number={2},
  pages={292},
  year={2024},
  publisher={IOP Publishing}
}

@article{friske2019more,
  title={More than just a wrinkle: a wave-like pattern in U g versus Lz from Gaia data},
  author={Friske, Jennifer KS and Sch{\"o}nrich, Ralph},
  journal={MNRAS},
  volume={490},
  number={4},
  pages={5414--5423},
  year={2019},
  publisher={Oxford University Press}
}

@ARTICLE{modak2025characterizing,
       author = {{Modak}, Shaunak and {Ostriker}, Eve C. and {Hamilton}, Chris and {Tremaine}, Scott},
        title = "{Characterizing Density and Gravitational Potential Fluctuations of the Interstellar Medium}",
      journal = {arXiv e-prints},
         year = 2025,
        month = jun,
          eid = {arXiv:2506.17387},
        pages = {arXiv:2506.17387},
archivePrefix = {arXiv}
}

@article{hunt2025milky,
  title={Milky Way dynamics in light of Gaia},
  author={Hunt, Jason AS and Vasiliev, Eugene},
  journal={New Astronomy Reviews},
  pages={101721},
  year={2025},
  publisher={Elsevier}
}

@article{hamilton2024kinetic,
  title={Kinetic theory of stellar systems: A tutorial},
  author={Hamilton, Chris and Fouvry, Jean-Baptiste},
  journal={PoP},
  volume={31},
  number={12},
  year={2024},
  publisher={AIP Publishing}
}

@BOOK{Binney2008-ou,
	edition = {2nd},
	title = {Galactic {Dynamics}},
	isbn = {0-691-13027-2},
	publisher = {Princeton University Press},
	author = {Binney, James and Tremaine, Scott},
	month = jan,
	year = {2008},
}

@article{foreman2013emcee,
  title={emcee: the MCMC hammer},
  author={Foreman-Mackey, Daniel and Hogg, David W and Lang, Dustin and Goodman, Jonathan},
  journal={PASP},
  volume={125},
  number={925},
  pages={306},
  year={2013},
  publisher={IOP Publishing}
}

@article{chiba2025origin,
  title={Origin of the two-armed vertical phase spiral in the inner Galactic disc},
  author={Chiba, Rimpei and Frankel, Neige and Hamilton, Chris},
  journal={MNRAS},
  volume={543},
  number={3},
  pages={2159--2179},
  year={2025},
  publisher={Oxford University Press}
}

@article{hamilton2026galactokinetics,
  title={Galactokinetics},
  author={Hamilton, Chris and Modak, Shaunak and Tremaine, Scott},
  journal={ApJ},
  volume={997},
  number={1},
  pages={28},
  year={2026},
  publisher={IOP Publishing}
}

@article{binney2024chemodynamical,
  title={Chemodynamical models of our Galaxy},
  author={Binney, James and Vasiliev, Eugene},
  journal={MNRAS},
  volume={527},
  number={2},
  pages={1915--1934},
  year={2024},
  publisher={Oxford University Press}
}

@ARTICLE{bland2019,
       author = {{Bland-Hawthorn}, Joss and {Sharma}, Sanjib and {Tepper-Garcia}, Thor and {Binney}, James and {Freeman}, Ken C. and {Hayden}, Michael R. and {Kos}, Janez and {De Silva}, Gayandhi M. and {Ellis}, Simon and {Lewis}, Geraint F. and {Asplund}, Martin and {Buder}, Sven and {Casey}, Andrew R. and {D'Orazi}, Valentina and {Duong}, Ly and {Khanna}, Shourya and {Lin}, Jane and {Lind}, Karin and {Martell}, Sarah L. and {Ness}, Melissa K. and {Simpson}, Jeffrey D. and {Zucker}, Daniel B. and {Zwitter}, Toma{\v{z}} and {Kafle}, Prajwal R. and {Quillen}, Alice C. and {Ting}, Yuan-Sen and {Wyse}, Rosemary F.~G.},
        title = "{The GALAH survey and Gaia DR2: dissecting the stellar disc's phase space by age, action, chemistry, and location}",
      journal = {\mnras},
         year = 2019,
        month = jun,
       volume = {486},
       number = {1},
        pages = {1167-1191},
 primaryClass = {astro-ph.GA}
}

@article{bland2021galactic,
  title={Galactic seismology: the evolving ‘phase spiral’after the Sagittarius dwarf impact},
  author={Bland-Hawthorn, Joss and Tepper-Garc{\'\i}a, Thor},
  journal={MNRAS},
  volume={504},
  number={3},
  pages={3168--3186},
  year={2021},
  publisher={Oxford University Press}
}
\bibliographystyle{aasjournal}

\end{document}